\documentclass[11pt,a4paper]{article}
\pdfoutput=1
\usepackage{jheppub}
\usepackage{amsfonts}
\usepackage{bbm}
\usepackage{youngtab}
\usepackage{rotfloat}

\def\a{\alpha}
\def\b{\beta}
\def\c{\gamma}

\newcommand{\bC}{\ensuremath{\mathbb{C}}}

\newcommand{\bZ}{\ensuremath{\mathbb{Z}}}

\newcommand{\scH}{\ensuremath{\mathcal{H}}}

\newcommand{\scP}{\ensuremath{\mathcal{P}}}

\newcommand{\scS}{\ensuremath{\mathcal{S}}}

\newcommand{\fraksl}{\ensuremath{\mathfrak{sl}}}

\newcommand{\cC}{\mathcal{C}}

\newcommand{\cW}{\mathcal{W}}

\newcommand{\Tr}{\mbox{Tr}}

\newcommand{\Li}{{\rm Li}}

\def\bea{\begin{eqnarray}}
\def\eea{\end{eqnarray}}
\def\be{\begin{equation}}
\def\ee{\end{equation}}
\def\ba{\begin{align}}
\def\ea{\end{align}}

\newcommand{\bem}{\begin{pmatrix}}
\newcommand{\eem}{\end{pmatrix}}

\def\={\;  = \;}
\def\+{\, + \,}

\def\bar{\overline}

\def\rt2{\sqrt{2}}

\preprint{}

\title{Super-A-polynomials for Twist Knots}

\author{Satoshi Nawata${}^1$, P. Ramadevi${}^2$, Zodinmawia${}^2$\\
\hspace{8cm} {\it with a program by} Xinyu Sun${}^3$}
\affiliation{${}^1$Perimeter Institute for Theoretical Physics \\
Waterloo, Ontario, N2L 2Y5, Canada\vspace{.2cm}}

\affiliation{${}^2$Department of Physics, Indian Institute of Technology Bombay,\\
 Mumbai, India, 400076\vspace{.2cm}}
 \affiliation{${}^3$Department of Mathematics, Xavier University of Louisiana,\\
 New Orleans, LA, USA 70125\vspace{.2cm}}
\emailAdd{snawata@gmail.com, ramadevi@phy.iitb.ac.in, zodin@phy.iitb.ac.in, xsun@xula.edu}

\abstract{We conjecture formulae of the colored superpolynomials for a class
of twist knots $K_p$ where $p$ denotes the number of full twists. The validity of the formulae is checked by applying differentials and taking special limits. Using the formulae, we compute both the classical and quantum super-$A$-polynomials for the twist knots with small values of $p$. The results support the categorified versions of the generalized volume conjecture and the quantum volume conjecture. Furthermore, we obtain the evidence that the $Q$-deformed $A$-polynomials can be identified with the augmentation polynomials of knot contact homology in the case of the twist knots.
}


\begin{document}
\Yboxdim4pt

\maketitle
\section{Introduction}\label{sec:intro}

\noindent \emph{\bf Historical Background}

One of the challenging problems in knot theory is the 
classification of knots and links. 
Alexander polynomial $\Delta(K;z)$ \cite{Alexander:1928}, where $K$ represents the knot and $z$ denotes the polynomial variable, is the 
first polynomial invariant  obtained using a skein 
relation which partially attempted 
the classification problem. However, the Alexander polynomial 
is zero for all unlinked knots. Moreover, the polynomials are the same for the knot  $K$ from its mirror image $K^*$. 
Another skein relation introduced by Jones~\cite{Jones:1985dw}
revealed that the Jones polynomial $J(K;q)$ 
can distinguish, in general, a chiral knot $K$ from its mirror  $K^*$.  In addition, unlinked knots have non-trivial Jones polynomials 
suggesting that $J(K;q)$ is more powerful than $\Delta(K;z)$.
Even though further two-variable generalizations, called HOMFLY(-PT) polynomials $P(K;a,q)$  \cite{Freyd:1985dx,PT} and Kauffman polynomials $F(K;a,q)$ \cite{Kauffman:1987,Kauffman:1990}, were found, none of these polynomials could solve the
classification problem.

The seminal work by Witten \cite{Witten:1988hf} demonstrated that 
three-dimensional Chern-Simons theory with a compact gauge group  $G$ provides a natural framework for 
the study of knots and links. Particularly, the expectation value of
the Wilson loop observable along a knot $K$ gives the polynomial invariant $V_R^G(K;q)$ 
of the knot $K$. The suffix $R$ denotes the representation of the 
group $G$ and the polynomial variable $q=\exp[2 \pi i/(k+C_v)]$ where 
$k\in \bZ$ is the Chern-Simons level and $C_v$ is the quadratic Casimir in the adjoint representation.
For a class of $(n-1)$-th rank symmetric representation ($R=\overbrace{\yng(2)\cdots\yng(2)}^{n-1}\equiv \scS^{n-1}$) of $SU(2)$, the field theoretic invariant denoted by $V_n^{SU(2)}$ turns out to be  proportional to
the colored Jones polynomial
\begin{equation}\label{jones}
{V_{n}^{SU(2)}(K;q) \over V_{n}^{SU(2)}(\bigcirc;q)}=
J_{n}(K;q)~,
\end{equation}
where we denote the unknot by $\bigcirc$. In the case of the fundamental representation $n=2$ (or $R=\yng(1)$), this reduces to the original Jones polynomial, {\it i.e.} $J_{n=2}(K;q)=J(K;q)$.
Similarly, colored HOMFLY polynomials\footnote{To fix the notation, in this paper, we use the skein relation $a^{1/2}P_{L_+}-a^{-1/2}P_{L_-}=(q^{1/2}-q^{-1/2})P_{L_0}$ for HOMFLY polynomials $P(K;a,q)=P_{n=2}(K;a,q)$.}
are related to $SU(N)$ Chern-Simons invariants as
\begin{equation}\label{HOMFLY}
{V_{n}^{SU(N)}(K;q) \over V_{n}^{SU(N)}(\bigcirc;q)}=P_{n}(K;a=q^N,q)~.
\end{equation}
For $SO(N)$ gauge group, the Chern-Simons invariant in the $N$-dimensional vector representation
gives the Kauffman polynomial $F(K;a,q)$.
Besides giving the known Jones, HOMFLY and Kauffman polynomials,
the paper \cite{Witten:1988hf} paved the way to explore a large family of new invariants for knots carrying arbitrary representations of any compact gauge group \cite{RamaDevi:1992dh}, which improved classification \cite{Ramadevi:1993hu}.
 \vspace{0.2cm}

\noindent \emph{\bf Categorifications} 

Explicit evaluation of  colored HOMFLY
polynomials for many knots gives Laurent series expansion:
\begin{equation}
P_R(a,q)= \sum_{i,j} c_{i,j} q^i a^j~,
\end{equation}
where $c_{i,j}$ coefficients are integers, which suggests that there must be an underlying topological interpretation for these integers coefficients. After the pioneering works by Khovanov \cite{Khovanov:2000,Khovanov:2003}, the bi-graded homology theory called the \emph{colored $\fraksl_2$ knot homology} $\scH_{i,j}^{\fraksl_2,R}$  was introduced \cite{Webster:2010,Cooper:2010,Frenkel:2010} as a categorification of the 
colored Jones polynomial. We denote the Poincar\'e polynomial of the colored $\fraksl_2$ homology $\scH_{i,j}^{\fraksl_2,R}$ by
\begin{equation}\label{Khovanov}
\scP^{\fraksl_2}_R(K;q,t)= \sum_{i,j} t^j q^i \dim {\cal H}_{i,j}^{\fraksl_2, R}(K)~,
\end{equation} 
so that the subscripts $i$ and $j$ are called the quantum (polynomial) grading and the homological gradings respectively.
The $q$-graded Euler
characteristic of the colored $\fraksl_2$ knot homology gives the colored Jones polynomial:
\begin{equation}
J_R(K;q)=\scP^{\fraksl_2}_R(q,t=-1)=\sum_{i,j} (-1)^j q^i  \dim {\cal H}_{i,j}^{\fraksl_2,R}(K)~.
\end{equation}
Hence, 
it is clear from this point of view that  coefficients of Jones can be interpreted as 
 dimensions of vector spaces of homological theory. The case of  higher rank gauge groups has been also studied in the context of bi-graded homology theory, which brought about the \emph{colored $\fraksl_N$ homology} $\scH_{i,j}^{\fraksl_N,R}$ \cite{Khovanov:2004,Mackaay:2007,Mazorchuk:2007,Wu:2009}. The Poincar\'e polynomial of the colored $\fraksl_N$ homology  
\begin{eqnarray}
\scP^{\fraksl_N}_R(K;q,t)&=& \sum_{j,i} t^i q^j \dim \scH_{i,j}^{\fraksl_N,R}(K)~.
\end{eqnarray}
is related to the colored HOMFLY polynomial via
\be
\scP^{\fraksl_N}_R(K;q,t=-1)=P_R(K;a=q^N,q) \ .
\ee

 \begin{table}
\begin{center}
\begin{tabular}{|c|c|c|c||}
\hline 
\textbf{Group} &   \textbf{Knot invariants} &\textbf{Quantum operators} \tabularnewline\hline 
\hline \rule{0pt}{5mm}
$SU(2)$ &  \footnotesize{\rm{colored Jones} $J_n(K;q)$} & \footnotesize{$\widehat{A}(K;\hat{x}, \hat{y};q)J_n(K;q)=0$} \tabularnewline
\hline \rule{0pt}{5mm}
$SU(N)$ &  \footnotesize{ \rm{colored HOMFLY} $P_n(K;a,q)$} &\footnotesize{ $\widehat{A}^{Q}(K;\hat{x}, \hat{y};a,q)P_n(K;a,q)=0$} \tabularnewline
\hline
\hline
&$\downarrow$ \ \textbf{Refinement} &$\downarrow$ \ \textbf{Refinement}\tabularnewline\hline 
\hline \rule{0pt}{5mm}
$SU(2)$ &\footnotesize{ colored $\fraksl_2$ homological inv. $\scP^{\fraksl_2}_n(K;q,t)$ }&\footnotesize{ $\widehat{A}^{\rm ref}_{N=2}(K;\hat{x}, \hat{y};q,t)\scP^{\fraksl_2}_n(K;q,t)=0$} \tabularnewline
\hline \rule{0pt}{5mm}
$SU(N)$ &\footnotesize{ colored $\fraksl_N$ homological inv. $\scP^{\fraksl_N}_n(K;q,t)$} & \footnotesize{$\widehat{A}^{\rm ref}_{N}(K;\hat{x}, \hat{y};q,t)\scP^{\fraksl_N}_n(K;q,t)=0$} \tabularnewline
\hline\hline
&$\downarrow$ \ \textbf{Unification} &$\downarrow$ \textbf{Unification} \ \tabularnewline\hline 
\hline
$SU(N)$ & \footnotesize{ \rm{colored superpolynomial} $\scP_n(K;a,q,t)$} & \footnotesize{$\widehat{A}^{\rm super}(K;\hat{x}, \hat{y};a,q,t)\scP_n(K;a,q,t)=0$} \tabularnewline
\hline
\end{tabular}\caption{Knot invariants and categorifications of quantum volume conjecture. We place the knot  invariants with the symmetric representation $\scS^{n-1}$ of special unitary groups  in historical order.  On the right, the corresponding quantum volume conjectures are written.  The action of the operators $\hat x$ and $\hat y$ are as in \eqref{xyactionJ}. }
\label{tab:notation}
\end{center}
\end{table}

Furthermore, a categorifaction of colored HOMFLY polynomials $P_R(K;a,q)$ as polynomials with two variables $(a,q)$ has led to the triply-graded homology theory called the \emph{colored HOMFLY homology} $\scH_{i,j,k}^{R}$ \cite{Khovanov:2005a,Dunfield:2005si,Khovanov:2005b,Mackaay:2008,Webster:2009} whose Poincar\'e polynomial is called the \emph{colored superpolynomial}
\bea
\scP_R(K;a,q,t)= \sum_{i,j,k} a^i q^j t^k\dim \scH_{i,j,k}^R(K) \ .
\eea
The $(a,q)$-graded Euler characteristic of the triply-graded homology theory is equivalent to the colored HOMFLY polynomial
\be
P_R(K;a,q)= \sum_{i,j,k} (-1)^k a^i q^j \dim \scH_{i,j,k}^R(K) \ .
\ee
It is important to stress that in this paper we consider \emph{reduced} homological knot invariants, {\it i.e.} $\scP_R(\bigcirc)=\scP^{\fraksl_N}_R(\bigcirc)=\scP^{\fraksl_2}_R(\bigcirc)=1$. 

 \begin{figure}[b]
 \centering
    \includegraphics[width=15cm]{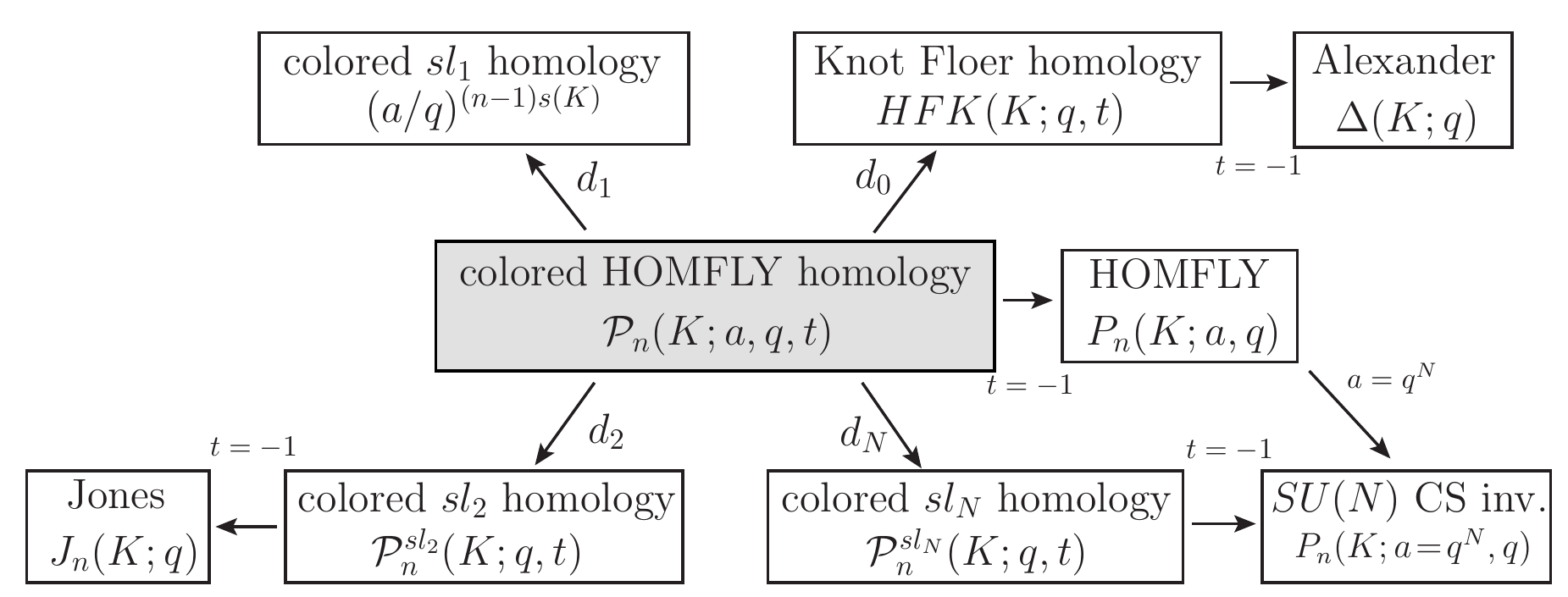}
    \caption{Schematic diagram for relations of various knot invariants. The bi-graded homologies are obtained by acting differentials on the colored HOMFLY homology. For the colored $\fraksl_{N\ge2}$ homological invariants, the relation is written in \eqref{remainder}. For any knot $K$, a colored $\fraksl_1$ homology is one-dimensional such that  $R^{\fraksl_1}_{n} (K;a,q,t)=(a/q)^{(n-1)s(K)}$ where $s(K)$ is called the $s$-invariant of the knot $K$.    The knot Floer homology is indeed the $\mathfrak{gl}(1|1)$ homology whose Poincar\'e polynomial  $R^{\fraksl_0}_{n=2} (K;a=1,q,t)=HFK(K;q,t)$  is the categorification of the Alexander polynomial $\Delta(K;q)$.   }
    \label{fig:chart}
 \end{figure}

Naively thinking, one might conclude that the colored superpolynomial is the same as the Poincar\'e polynomial of the colored $\fraksl_N$ homology. However, in general, this is \emph{not} the case $\scP_R(K;a=q^N,q,t)\neq \scP^{\fraksl_N}_R(K;q,t)$ as emphasized in \cite{Dunfield:2005si,Gukov:2011ry,Fuji:2012pm,Fuji:2012nx}. Rather, they are related by differentials
\be
\left(\scH_{*,*,*}^{R},d_N\right)\cong \scH_{*,*}^{\fraksl_N,R} \quad\quad {\rm for}\  N\in \bZ\ .
\ee
where we drop off triply-graded homology groups $d_N:\scH_{i,j,k}\to \scH_{i+\a,j+\b,k+\c}$ paired by the differential $d_N$ of $(a,q,t)$-degree $(\a,\b,\c)$  as being exact. 
As Poincar\'e polynomials, this statement can be summarized that the colored superpolynomial has the structure:
\be
\scP_R (K;a,q,t)  = 
R^{\fraksl_N}_R (K;a,q,t) + (1 + a^{\a} q^{\b} t^{\c}) Q^{\fraksl_N}_R (K;a,q,t) \,,
\label{superright}
\ee
where the part $Q^{\fraksl_N}_R$ corresponds to the homologies annihilated by the differential $d_N$ and a bi-graded homological knot invariant is a specialization of the ``remainder'':
\be\label{remainder}
R^{\fraksl_N}_R (K;a=q^N,q,t) =  \scP^{\fraksl_N}_{R} (K;q,t) \quad {\rm for} \quad N\ge2 \,,
\ee
Note that $R^{\fraksl_N}_R (K;a,q,t)$ and $Q^{\fraksl_N}_R (K;a,q,t)$ are polynomials with non-negative coefficients.
Since bi-graded homologies can be obtained from the colored HOMFLY homology by acting differentials, the colored superpolynomial as the Poincar\'e polynomial of the colored HOMFLY homology is considered as an unified knot invariant \cite{Dunfield:2005si}. (See Figure \ref{fig:chart} for relations of various knot invariants.) 

For a $\scH$-thick knot $K$ (see \cite{Khovanov:2002,Dunfield:2005si} for the definition) such as the torus knot $T^{3,4}={\bf 8_{19}}$ and $\bf 9_{42}$, the part $Q^{\fraksl_N}_R (K;a,q,t)$  is not necessarily zero. On the other hand, for a $\scH$-thin knot $K$ such a 2-bridge knot, we always have $Q^{\fraksl_N}_R (K;a,q,t)=0$ for $N\ge 2$, {\it i.e.} $\scP_R(K;a=q^N,q,t)= \scP^{\fraksl_N}_R(K;q,t)$ for $N\ge 2$. At this juncture, we would  clarify  that the twist knots considered in this paper are thin knots since they belong to a class of the 2-bridge knots.

% The categorification of Kauffman polynomials had also been considered in 
% \cite{Gukov:2005qp,Khovanov:2007}. The homological invariants are usually more powerful than the knot invariants of Jones type \cite{Ozsvath:2003,Baldwin:2006,Kronheimer:2010}.

\vspace{0.2cm}

\noindent \emph{\bf Topological strings and refined Chern-Simons theory}

In a parallel development, the knot invariants have also been studied in the context of string theory. The first step was made in \cite{Witten:1992fb} in which Chern-Simons theory on a three-manifold $M$ is realized in the open topological $A$-model on $T^*M$ where Lagrangian branes wrap the zero section $M$ of $T^*M$. Using the fact that the topological $A$-model can be embedded into Type IIA string theory, it was shown in \cite{Gopakumar:1998ki,Gopakumar:1998ii,Gopakumar:1998jq} that Chern-Simons partition functions are related to D2-D0 bound states (Gopakumar-Vafa invariants) via geometric transitions. Furthermore, by adding other Lagrangian branes to the setting, the expectation value of Chern-Simons Wilson loop operators in the fundamental representation can be reformulated into the generating function of D4-D2 bound states \cite{Ooguri:1999bv}. These works and further developments \cite{Ramadevi:2000gq,Labastida:2000yw,Labastida:2001ts} have revealed that  geometric transitions are very powerful that the knot invariants are expressed in terms of cohomology groups of moduli spaces of BPS states. Moreover, the one-parameter deformation $t$ of the colored $\fraksl_2$ homological knot invariants \eqref{Khovanov} was also interpreted in topological strings as the extra fugacity of the index which counts D4-D2 bound states \cite{gukov:2004hz}. This work as well as the equivariant instanton partition function of Nekrasov \cite{Nekrasov:2002qd} led to the study of refined topological strings \cite{Iqbal:2007ii,Awata:2008ed}.\footnote{We refer the reader to \cite{Gukov:2007tf,Taki:2008hb,Awata:2009sz,Iqbal:2011kq} for further developments, and to \cite{Witten:2011zz} which gives another 
viewpoint of the homological grading from five-brane world volume theory.}

From the perspective of refined topological strings,
refined Chern-Simons theory was formulated in \cite{Aganagic:2011sg,Aganagic:2012ne}, in which
one-parameter deformations of the modular
transformation matrices $S$ and $T$ were especially proposed. 
This enabled the evaluation of refined torus knot
invariants, by using knot operators, which involve Macdonald polynomials. After suitable change of variables,
refined torus knot invariants carrying fundamental representation of $SU(N)$ was shown to agree with the Poincar\'e polynomials of $\fraksl_N$ homology \cite{Aganagic:2011sg}. Recently, the colored superpolynomials of torus knots $T^{2,2p+1}$ of the type $(2,2p+1)$ 
have been computed by using braiding operations in refined Chern-Simons theory \cite{Fuji:2012pm}.

For non-torus knots, the same methods cannot be applied to the explicit calculations of the colored superpolynomials. In fact, even colored HOMFLY polynomials $P_n(K;a,q)$ for non-torus knots $K$ are generally not known for $n>3$ since the information about quantum $SU(N)$ Racah coefficients ({\it a.k.a.} quantum $6j$-symbols for $U_q(\fraksl_N)$) is lacking to date. 
 However, for the figure-eight knot ${\bf 4_1}$ which is the simplest non-torus knot, the forms of the colored HOMFLY polynomials $P_n({\bf 4_1};a,q)$ were conjectured in \cite{Itoyama:2012fq} by looking at the pattern of the expressions for $n=2,3,4$. In addition, the results were  refined from the colored HOMFLY polynomials $P_n({\bf 4_1};a,q)$ to the colored superpolynomials $\scP_n({\bf 4_1};a,q,t)$.  Nevertheless, the knowledge of the colored superpolynomials for other non-torus knots are still very limited.

\vspace{0.2cm}

\noindent \emph{\bf Volume conjecture and $A$-polynomials}

For actual calculations of knot invariants such as \eqref{jones} and \eqref{HOMFLY}, purely algebraic operations, such as quantum groups and representations of braid groups, are indeed involved. Even though knots are located in three-manifolds, it is not clear from their definitions that they are related to topology of three-manifolds. The most promising candidate which provides connection between ``quantum invariants'' of knots and ``classical'' three-dimensional topology is the \emph{volume conjecture}. 
With the physical insight \cite{Witten:1988hc} behind, Kashaev observed in \cite{Kashaev:1996kc} that the asymptotic behavior of his knot invariant of
any hyperbolic knot, which 
is later identified with the specific value of the colored Jones polynomial $J_n(K;q=e^{2\pi i\over n})$,
describes the hyperbolic volume of its knot complement. The statement of the volume conjecture is summarized in the following form \cite{Murakami:1999}:
\begin{equation}\label{volumeconj}
\lim_{n \rightarrow \infty} {2 \pi \over n}{\rm log}\Big|J_n(K;q=e^{2\pi i \over n})\Big|= {\rm Vol}(S^3\backslash K)~.
\end{equation}
The volume conjecture is further generalized by incorporating yet another knot invariant, called \emph{$A$-polynomial} \cite{Cooper:1994} which is a character variety of $SL(2,\bC)$-representation of the fundamental group of the knot complement. More precisely, it is conjectured in  \cite{Gukov:2003na} that, taking double scaling limit, the equation
\begin{equation}\label{volumeconj2}
\log y=-x{d \over d x} \left[ \lim_{\substack{n,k \rightarrow \infty\\ e^{i \pi n/k}=x}}\
{1 \over k} \log J_n(K;q=e^{2 \pi i\over k})\right]~,
\end{equation}
gives the zero locus of the $A$-polynomial $A(K;x,y)$ of the knot $K$. 

Physically, it is natural to quantize the $A$-polynomial $A(K;x,y)$, which result in the operator ${\widehat A}(K;\hat x, \hat y; q)$.  Taking $q=e^{\hbar}=1$ gives the classical $A$-polynomial $A(K;x,y)$. Then, the quantum version of the volume conjecture can be stated that the quantum $A$-polynomial annihilates the Chern-Simons partition function, {\it i.e.} ${\widehat A} Z_{CS}=0$ \cite{Gukov:2003na}. On the mathematics side, this conjecture which is called the \emph{AJ conjecture}\footnote{We use the names ``the AJ conjecture'' and ``the quantum volume conjecture''
interchangeably in the rest of this paper.} \cite{Garoufalidis:2003a,Garoufalidis:2003b} was more concretely expressed 
that 
\be\label{AJ}
\widehat{A}(K;\hat x,\hat y,q) J_n(K;q) =0~
\ee
where the operators $\hat x$ and  $\hat y$ acts on the set of the colored Jones polynomials as 
\be\label{xyactionJ}
\hat x J_n(K,q^n)=q^n J_n(K,q^n) \ , \quad \hat y J_n(K;q)=J_{n+1}(K;q) \ .
\ee
Therefore,  the difference equation of the colored Jones polynomials of minimal order
\be
b_k(\hat x,q)J_{n+k}(K;q)+\cdots+b_0(\hat x,q)J_{n}(K;q)=0
\ee
amounts to  the quantum $A$-polynomial ${\widehat A}(K;\hat x, \hat y; q)=\sum_{j=0}^k b_j(\hat x,q)\hat y^j$.

Furthermore, the conjectures \eqref{volumeconj2} and \eqref{AJ} were categorified in \cite{Fuji:2012pm,Fuji:2012nx} for colored superpolynomials by incorporating the two-parameter $(a,t)$. Namely, the difference equation of the colored superpolynomials $\scP_n(K;a,q,t)$ of minimal order gives the quantum \emph{super-$A$-polynomial} ${\widehat A}^{\rm super}(K;\hat{x},\hat{y};a,q,t)$, and, for $q=1$, it reduces to classical super-$A$-polynomial $A^{\rm super}(K;\hat{x},\hat{y};a,t)$ which can be obtained by substituting $\scP_n(K;a,q,t)$ for $J_n(K;q)$ in \eqref{volumeconj2}. Since the colored superpolynomials of the torus knots $T^{2,2p+1}$ and the figure-eight ${\bf 4_1}$ are available, the two-parameter deformations of the conjectures are explicitly checked in \cite{Fuji:2012pm,Fuji:2012nx}. The categorifications of the quantum volume conjecture are tabulated in Table~\ref{tab:notation}.
 \begin{figure}
\centering
    \includegraphics[width=5cm]{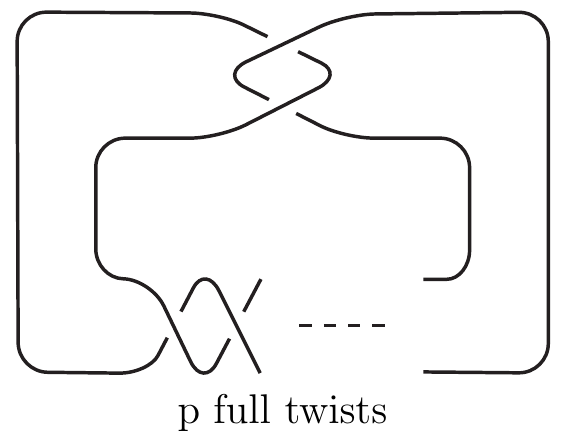}
    \caption{Twist knots $K_p$ with $p$ full twists}
    \label{fig:twistknot}
  \end{figure}

\vspace{0.2cm}

\noindent \emph{\bf Plan of the paper}

In this paper, we study a class of knots
called twist knots $K_p$ drawn in Figure \ref{fig:twistknot} where $p$  counts 
the number of right-handed full-twists $p>0$
or number of left-handed full-twists $p<0$. Since the twist knots $K_{p\neq 1}$ are all hyperbolic knots except the trefoil $K_{p=1}$, techniques to compute the colored superpolynomial $\scP_n(K_p;a,q,t)$ of the twist knot $K_p$ are not known. However, for the twist knots, there are data of several kinds available which suggest the form of the colored superpolynomial $\scP_n(K_p;a,q,t)$. The colored Jones polynomial of $K_{p>0}$ (of $K_{p<0}$) \cite{Masbaum:2003} have the colored Jones polynomial of the trefoil $K_1={\bf 3_1}$ (of the figure-eight $K_{-1}={\bf 4_1}$) as the main body and the twisting factors come along with it for the additional $p-1$ twists. Assuming that the colored superpolynomials of the twist knots have a similar structure, the main body gets replaced by the colored superpolynomial of the trefoil ${\bf 3_1}$ (the figure-eight $\bf 4_1$) for $K_{p>0}$ ($K_{p<0}$) which is already obtained \cite{Fuji:2012pm,Itoyama:2012fq}.
In addition, the colored superpolynomials $\scP_n(a,q,t)$ of the knots $K_2={\bf 5_2}$ and $K_{-2}={\bf 6_1}$ were computed up to $n=3$ \cite{Dunfield:2005si,Gukov:2011ry}. The results of these two knots indicate  the form of the twisting factors by which we propose the expressions of the colored superpolynomials $\scP(K_p;a,q,t)$. Hence, our strategy is similar to \cite{Itoyama:2012fq} in the sense that the colored superpolynomials are conjectured by observing the patterns of the available data.

The organization of the paper is as follows.
In \S\ref{sec:superpoly}, we conjecture
the formula for the colored superpolynomial of the twist knot $K_p$. We then apply differentials to the colored superpolynomial $\scP(K_p;a,q,t)$ to check if it is consistent with the homological knot invariants known in the mathematics literature. In \S\ref{sec:super-a-poly}, we attempt to obtain both the classical and quantum super-$A$-polynomials of the twist knots. For this section, we follow the analysis done in \cite{Fuji:2012pm,Fuji:2012nx}. To get the difference equations of the colored superpolynomials, we use the program {\tt iSumq.txt} written  by Xinyu Sun. Besides, the investigation is carried out on the conjecture recently proposed by Aganagic and Vafa \cite{Aganagic:2012jb}. The section  \S\ref{sec:discussions}  is devoted
to conclusions and future directions. In the appendix \S\ref{sec:equivalence}, we show that the formulae of two kinds for the colored superpolynomials are equivalent by using the Bailey chains. In  \S\ref{sec:sp}, \S\ref{sec:10crossings} and \S\ref{sec:OV}, we check the validity of the colored superpolynomials by taking special limits. These checks reinforces  the correctness of the colored superpolynomials we propose. We put tables and figures in \S\ref{sec: toobig} that are too big for the
main text.   Finally, we should mention that the program {\tt iSumq.txt}  and all the expressions we obtain in this paper are linked on the arXiv page as ancillary files.

\section{Colored superpolynomials}\label{sec:superpoly}
In this section, we propose the formulae of the colored superpolynomials of the twist knots. The planar projection of the twist knot $K_p$ has $2|p| + 2$ crossings where $2|p|$ of which come from the full twists, and the other 2 come from the negative clasp. (See Figure \ref{fig:twistknot}.) For small values of $p$, the twist knots are identified with knots in the Rolfsen's table \cite{KnotAtlas}. (See Table \ref{tab:Rolfsen}.) 

\begin{table}\begin{center}
\begin{tabular}{|c||c|c|c|c|c|c|c|c|c|}
\hline 
$p$ & -4 & -3 & -2 & -1 & 0 & 1 & 2 & 3 & 4\tabularnewline
\hline
\hline 
knots & ${\bf 10_1}$ & ${\bf 8_1}$ & ${\bf 6_1}$ & ${\bf 4_1}$ &${\bf 0_1}$  & ${\bf 3_1}$ & ${\bf 5_2}$ & ${\bf 7_2}$ & ${\bf 9_2}$\tabularnewline
\hline
\end{tabular}
\caption{The correspondence between the twist numbers and the knots in the Rolfsen's table \cite{KnotAtlas}.}
\label{tab:Rolfsen}
\end{center}
\end{table}

\subsection{Colored Jones polynomials of twist knots}\label{sec:jones}
To begin with, let us first review the colored Jones polynomials of the twist knots. In trying to obtain the unified Witten-Reshitekhin-Turaev (WRT) invariants for integral homology three-spheres, Habiro introduced the cyclotomic expansions of the colored Jones polynomials of the trefoil $\bf 3_1$ and the figure-eight $\bf 4_1$ by means of quantum group $U_q(\fraksl_2)$ \cite{Habiro:2000,Habiro:2002}. Inspired by this result, Masbaum showed  the cyclotomic expansion of the colored Jones polynomial of the twist knot $K_p$ by skein theory \cite{Masbaum:2003}:
\begin{equation}\label{Masbaum}
J_n(K_p;q)=\sum_{k=0}^{\infty} \cC_{K_p}(k) 
       \frac {\{n-k \} \{n-k+1\} \cdots \{n+k\}} {\{n\}} \ ,
\end{equation}
where
\begin{equation}
\cC_{K_p}(k)
=(-1)^{k+1} {q_I}^{k(k+3)/2}\sum_{\ell=0}^k (-1)^\ell q^{\ell(\ell+1)p} \{2\ell+1\}
 \frac {\{k \}! }{\{k+\ell+1 \}! \{k-\ell\}!} \ . 
\end{equation}
Here we use the notations as
\begin{eqnarray}
& & \{n\}={q_I}^{n}-{q_I}^{-n}\ ,  \quad {q_I}^2=q \ , \quad\{n\}!=\{n\}\{n-1\}\cdots \{1\}   \ .
\end{eqnarray}
Using  $q$-Pochhammer symbols $(z;q)_{k}=\prod_{j=0}^{k-1} (1-zq^j)$,  this formula \eqref{Masbaum} can be rearranged
\begin{eqnarray}\label{jones twist 3}
J_{n}(K_p;q)&=&\sum_{k=0}^{\infty}\sum_{\ell=0}^k q^{k}(q^{1-n};q)_k(q^{n+1};q)_k\cr
& & \qquad \times 
       (-1)^\ell q^{\ell(\ell+1)p+\ell(\ell-1)/2}(1-q^{2\ell+1}) 
       \frac {(q;q)_k}{(q;q)_{k+\ell+1} (q;q)_{k-\ell}} \ .
\end{eqnarray} 
We should remark that, for the trefoil $K_1$ and the figure-eight $K_{-1}$, the summation over $\ell$ can be carried out so that the colored Jones polynomial can be written as a single sum \cite{Habiro:2002}
\bea
J_n(K_1;q)&=& \sum_{k=0}^\infty q^{k}\left(q^{1-n};q\right)_{k}(q^{1+n};q)_{k} \ ,\label{jones trefoil}\\
J_n(K_{-1};q)&=& \sum_{k=0}^\infty (-1)^{k} q^{-\frac{k(k+1)}{2}}(q^{1-n};q)_{k}(q^{1+n};q)_{k} \ . \label{jones f8}
\eea
For  the sake of later argument, let us bring
the double-sum formula \eqref{jones twist 3} into a multi-sum formula \cite{Habiro:2002,Hikami:2007}:
\bea\label{jones twist 1}
J_n(K_{p>0};q)=\sum^\infty_{s_p\ge\cdots\ge s_1\ge0} q^{s_p}\left(q^{1-n};q\right)_{s_p}\left(q^{1+n};q\right)_{s_p}\prod^{p-1}_{i=1}q^{s_i(s_i+1)}\left[\begin{array}{c}s_{i+1}\\ s_{i} \end{array}\right]_q
\eea
for $p>0$, and 
\bea\label{jones twist 2}
J_n(K_{p<0};q)=\sum^\infty_{s_{|p|}\ge\cdots\ge s_1\ge0}(-1)^{s_{|p|}} q^{-\frac{s_{|p|}(s_{|p|}+1)}{2}}\left(q^{1-n};q\right)_{s_{|p|}}\left(q^{1+n};q\right)_{s_{|p|}}\cr
\hspace{3cm}\times\prod^{|p|-1}_{i=1}q^{-s_i(s_{i+1}+1)}\left[\begin{array}{c}s_{i+1}\\ s_{i} \end{array}\right]_q
\eea
for $p<0$, where we use the $q$-binomial
\be
\left[\begin{array}{c}n\\ k \end{array}\right]_q\equiv\frac{(q;q)_n}{(q;q)_k(q;q)_{n-k}} \ .
\ee
The equivalence of the double-sum expression \eqref{jones twist 3} with the multi-sum ones \eqref{jones twist 1} and  \eqref{jones twist 2} will be shown in \S\ref{sec:equivalence} by the Bailey's lemma. Now, we can see that the formula \eqref{jones twist 1} (the formula\eqref{jones twist 2}) is built out of the colored Jones polynomial of the trefoil \eqref{jones trefoil} (the figure-eight \eqref{jones f8}) with $p-1$ additional twisting factors. In addition, the twisting factors in \eqref{jones twist 1} and \eqref{jones twist 2} turn out to be related via
\bea\label{product}
q^{s_i(s_i+1)}\left[\begin{array}{c}s_{i+1}\\ s_{i} \end{array}\right]_q\quad \longleftarrow (q\leftrightarrow q^{-1})\longrightarrow\quad q^{-s_i(s_{i+1}+1)}\left[\begin{array}{c}s_{i+1}\\ s_{i} \end{array}\right]_q \ .
\eea

\subsection{Colored superpolynomials of twist knots}
From the colored Jones polynomials, it is natural to expect that the colored superpolynomials of the twist knots also have a similar structure. Namely, they can be built out from the colored superpolynomials of the trefoil ${\bf 3_1}$ and the figure-eight ${\bf 4_1}$. Fortunately,
the expression of the colored superpolynomial of the trefoil $\bf 3_1$ was already obtained in \cite{Fuji:2012pm} by using refined Chern-Simons theory. Though the more concise formula was presented in (2.23) of \cite{Fuji:2012nx}, we rather use the different expression 
\bea
\scP_n (K_1;a,q,t)= (-t)^{-n+1} \sum_{k=0}^{\infty}q^k  \frac{(-a t q^{-1};q)_k}{(q;q)_k}  (q^{1-n};q)_k (-a t^3 q^{n-1};q)_k  \ ,
\label{Paqt31}
\eea
which is the generalization of  \eqref{jones twist 1}.
Besides, the colored superpolynomial of the figure-eight $\bf 4_1$ was recently proposed in \cite{Itoyama:2012fq} and it was transformed into the succinct form (See (2.12) of \cite{Fuji:2012nx}.):
\be
\scP_n (K_{-1};a,q,t) = \sum_{k=0}^{\infty} (-at^2)^{-k} q^{-k(k-3)/2} \frac{(-a t q^{-1};q)_k}{(q;q)_k}  (q^{1-n};q)_k (-a t^3 q^{n-1};q)_k \, .
\label{Paqt41}
\ee
It is clear that this is the generalization of \eqref{jones twist 2}. Furthermore, the colored superpolynomials of ${\bf 5_2}=K_2$ and ${\bf 6_1}=K_{-2}$ were obtained in \cite{Dunfield:2005si, Gukov:2011ry} up to $n=3$. The formulae are recapitulated in Table \ref{tab:52 61}. It  should be noted that we change variables in such a way that the colored superpolynomials in Table \ref{tab:52 61} reduces to the colored Jones polynomials \eqref{jones twist 3} when $a=q^2$ and $t=-1$. With these data, we shall do an educative guess on the form of the colored superpolynomial $\scP_n(K_p;a,q,t)$ of the twist knots $K_p$ which is the generalization of the form \eqref{jones twist 1} or \eqref{jones twist 2}.

\begin{table}
\begin{center}
\begin{tabular}{|c|p{12cm}|}
\hline 
$\scP_2({\bf 5_2};a,q,t)$ & $a q^{-1} + a t + a q t^2 + a^2 q^{-1} t^2
+ a^2 t^3 + a^2 q t^4 + a^3 t^5$\tabularnewline
\hline 
$\scP_3({\bf 5_2};a,q,t)$ &
$a^2q^{-2}+(a^2q^{-1}+a^2)t+(2a^2q+a^2q^2+a^3q^{-2}+a^3q^{-1})t^2 $\\[3pt]
\rule{0pt}{4mm}
&$+(a^2q^2+a^2q^3+2a^3+2a^3q)t^3+(a^2q^4+2a^3q+3a^3q^2+a^3q^3+a^4)t^4$\\[3pt]
\rule{0pt}{4mm}
&$ +(2a^3q^3+2a^3q^4+a^4+2a^4q+a^4q^2)t^5$\\[3pt]
\rule{0pt}{4mm}
&$ +(a^3q^4+a^3q^5+a^4q^2+3a^4q^3+a^4q^4)t^6$\\[3pt]
\rule{0pt}{4mm}
&$ +(a^4q^3+2a^4q^4+a^4q^5+a^5q^2+a^5q^3)t^7$\\[3pt]
\rule{0pt}{4mm}
&$+(a^4q^6+a^5q^3+a^5q^4)t^8+(a^5q^5+a^5q^6)t^9+a^6q^5t^{10}$
\tabularnewline
\hline \rule{0pt}{5mm}
$\scP_2({\bf 6_1};a,q,t)$ & $a^{-2} t^{-4}+a^{-1}  q^{-1}  t^{-3} +a^{-1} t^{-2} + q^{-1} t^{-1}+ a^{-1} q t^{-1} + 2 + q t  
+ a t^2$
\tabularnewline
\hline \rule{0pt}{5mm}
$\scP_3({\bf 6_1};a,q,t)$ & $a^{-4} q^{-4} t^{-8}+(a^{-3} q^{-5}+a^{-3} q^{-4}) t^{-7}+(a^{-2} q^{-5} +a^{-3} q^{-3} +a^{-3} q^{-2})t^{-6}$\\
\rule{0pt}{4mm}
&$ +(a^{-2} q^{-4} +2 a^{-2} q^{-3} +a^{-3} q^{-2} +a^{-2} q^{-2} +a^{-3} q^{-1} )t^{-5}$\\
\rule{0pt}{4mm}
&$+(a^{-2}+a^{-1} q^{-4}+a^{-2} q^{-3}+a^{-1} q^{-3}+3a^{-2} q^{-2}+2a^{-2} q^{-1})t^{-4}$\\
\rule{0pt}{4mm}
&$ +(2a^{-2}+2a^{-1} q^{-3}+3a^{-1} q^{-2}+a^{-2} q^{-1}+a^{-1} q^{-1}+qa^{-2})t^{-3}$\\
\rule{0pt}{4mm}
&$+(4a^{-1}+q^{-3}+a^{-1} q^{-2}+4a^{-1} q^{-1}+qa^{-2}+qa^{-1})t^{-2}$\\
\rule{0pt}{4mm}
&$ +(1+2a^{-1}+2q^{-2}+3q^{-1}+3 qa^{-1}+q^2a^{-1})t^{-1}$\\
\rule{0pt}{4mm}
&$ +\left(5+q^{-1}+2 q+a^{-1}q^2+a^{-1}q^3\right)+\left(a+q^{-1}a+2 q+3 q^2+q^3\right) t$\\
\rule{0pt}{4mm}
&$ +\left(a+2 a q+a q^2+q^3\right) t^2+\left(a q^2+a q^3\right) t^3+a^2 q^2 t^4$
\tabularnewline
\hline
\end{tabular}
\caption{Colored superpolynomials of the knots ${\bf 5_2}$ and ${\bf 6_1}$: The superpolynomials ($n=2$) are drawn from Table 5.7 in \cite{Dunfield:2005si} where we change variables as $a^2\to a$, $q^2 \to q$. The colored superpolynomials for $n=3$ are pulled from Table 1 and 2 in \cite{Gukov:2011ry}. Note that we use the expressions of the mirror image of ${\bf 6_1}$. Namely, we make a change of variables as $a\to a^{-1}, q\to q^{-1}, t\to t^{-1}$ for ${\bf 6_1}$.}
\label{tab:52 61}
\end{center}
\end{table}

First, let us define the summands in \eqref{Paqt31} and \eqref{Paqt41} such that 
\bea
F_{n,k}(a,q,t)&\equiv&(-t)^{-n+1}q^k  \frac{(-a t q^{-1};q)_k}{(q;q)_k}  (q^{1-n};q)_k (-a t^3 q^{n-1};q)_k \cr
G_{n,k}(a,q,t)&\equiv& (-at^2)^{-k} q^{-k(k-3)/2} \frac{(-a t q^{-1};q)_k}{(q;q)_k}  (q^{1-n};q)_k (-a t^3 q^{n-1};q)_k 
\eea 
where they are related to each other via
\be
F_{n,k}(a,q,t)=  (-t)^{-n+1} (-at^2)^k q^{k(k-1)/2} G_{n,k}(a,q,t) \ .
\ee
From the structure of the colored Jones polynomials \eqref{jones twist 1} and \eqref{jones twist 2}, we suppose that the colored superpolynomials of the knots ${\bf 5_2}$ and ${\bf 6_1}$ encode $F_{n,k}$ and $G_{n,k}$ respectively as building blocks.  It is a straightforward, though tedious, calculation to show that the colored superpolynomials in Table \ref{tab:52 61} have the following structure:
\bea\label{evidence 52}
\scP_{2}({\bf 5_2};a,q,t) 
&=&F_{2,0}(a,q,t)+(1 + a t^2)F_{2,1}(a,q,t)\,,\cr
\scP_{3}({\bf 5_2};a,q,t)
&=&   F_{3,0}(a,q,t) + (1 + a t^2) F_{3,1}(a,q,t)+ (1 + a t^2 (1 + q) + a^2  t^4q^2)F_{3,2}(a,q,t)\ ,\cr 
&&
\eea
and
\bea\label{evidence 61}
\scP_{2}({\bf 6_1};a,q,t)&=&G_{2,0}(a,q,t)+(1 + a^{-1} t^{-2})G_{2,1}(a,q,t)\,,\cr
\scP_{3}({\bf 6_1};a,q,t)
&=&  G_{3,0}(a,q,t) +(1 + a^{-1} t^{-2}) G_{3,1}(a,q,t)\cr
&&\hspace{1.5cm}+(1 + a^{-1} t^{-2} (1 + q^{-1}) +a^{-2}t^{-4} q^{-2} )G_{3,2}(a,q,t) \ .
\eea
Then, it is plausible to think that the factors appear in \eqref{evidence 52} and \eqref{evidence 61} with $F_{n,k}$ and $G_{n,k}$ play the role of the twisting factors as in  the colored Jones polynomials.
Comparing \eqref{evidence 52} and \eqref{evidence 61} with the forms of \eqref{jones twist 1} and \eqref{jones twist 2}, it is natural to guess that
\bea
\scP_n({\bf 5_2};a,q,t)&=&\sum_{s_2\ge s_1\ge0}^{n-1} F_{n,s_2} (a,q,t)\ (at^2)^{s_1} q^{s_1(s_1-1)}\left[\begin{array}{c}s_{2}\\ s_{1} \end{array}\right]_q \ ,\cr
\scP_n({\bf 6_1};a,q,t)&=&\sum_{s_2\ge s_1\ge0}^{n-1} G_{n,s_2}(a,q,t)\  (at^2)^{-s_1} q^{-s_1(s_2-1)}\left[\begin{array}{c}s_{2}\\ s_{1} \end{array}\right]_q \ .
\eea
Furthermore, we conjecture the following.

\subsubsection*{Conjecture} 
\fbox{
\begin{minipage}[c]{15cm}
The colored superpolynomial of the twist knot $K_{p>0}$ is written as  
\bea\label{superpoly1}
\scP_{n}(K_{p>0};a,q,t)&=& (-t)^{-n+1} \sum^\infty_{s_p\ge\cdots\ge s_1\ge0} q^{s_p}  \frac{(-a t q^{-1};q)_{s_p}}{(q;q)_{s_p}}  (q^{1-n};q)_{s_p} (-a t^3 q^{n-1};q)_{s_p} \cr
&&\hspace{5cm}\times\prod^{p-1}_{i=1}(at^2)^{s_i} q^{s_i(s_i-1)}\left[\begin{array}{c}s_{i+1}\\ s_{i} \end{array}\right]_q
\eea
and, that of  the twist knot $K_{p<0}$ is
\bea\label{superpoly2}
&&\scP_{n}(K_{p<0};a,q,t)\cr
&&=\sum^\infty_{s_{|p|}\ge\cdots\ge s_1\ge0} (-at^2)^{-s_{|p|}} q^{-s_{|p|}(s_{|p|}-3)/2} \frac{(-a t q^{-1};q)_{s_{|p|}}}{(q;q)_{s_{|p|}}}  (q^{1-n};q)_{s_{|p|}} (-a t^3 q^{n-1};q)_{s_{|p|}} \cr
&& \hspace{5cm}\times\prod^{|p|-1}_{i=1}(at^2)^{-s_i} q^{-s_i(s_{i+1}-1)}\left[\begin{array}{c}s_{i+1}\\ s_{i} \end{array}\right]_q \ .
\eea
\end{minipage}
}
\\
\\
As in \eqref{product}, we find that 
\bea\label{inverse}
(at^2)^{s_i} q^{s_i(s_i-1)}\left[\begin{array}{c}s_{i+1}\\ s_{i} \end{array}\right]_q\longleftarrow(a,t,q \leftrightarrow a^{-1},q^{-1},t^{-1})\longrightarrow \quad  (at^2)^{-s_i} q^{-s_i(s_{i+1}-1)}\left[\begin{array}{c}s_{i+1}\\ s_{i} \end{array}\right]_q \nonumber\\
\eea
These formulae are the generalizations of \eqref{jones twist 1} and \eqref{jones twist 2}. To calculate the super-$A$-polynomials, it is convenient to find the double-sum expressions as in  \eqref{jones twist 3}. 
\\
\\
\fbox{
\begin{minipage}[c]{15cm}
For $K_{p>0}$, the double-sum expression is given by 
\bea\label{superpoly3}
\scP_{n}(K_{p>0};a,q,t)&=&(-t)^{-n+1} \sum^\infty_{k=0}  \sum_{\ell=0}^k q^{k} \frac{(-a t q^{-1};q)_{k}}{(q;q)_{k}}  (q^{1-n};q)_{k} (-a t^3 q^{n-1};q)_{k} \cr
&&\hspace{1cm}\times(-1)^\ell (a t^{2})^{p\ell} q^{(p+1/2)\ell(\ell-1)} \frac{1-at^2 q^{2\ell-1}}{(at^2q^{\ell-1};q)_{k+1}} \left[ \begin{array}{c} k \\ \ell \end{array}\right]_q   ,
\eea
and, for $K_{p<0}$, it is expressed by
\bea\label{superpoly4}
\scP_{n}(K_{p<0};a,q,t)&=& \sum^\infty_{k=0}  \sum_{\ell=0}^k q^{k} \frac{(-a t q^{-1};q)_{k}}{(q;q)_{k}}  (q^{1-n};q)_{k} (-a t^3 q^{n-1};q)_{k} \cr
&&\hspace{1cm}\times(-1)^\ell (a t^{2})^{p\ell} q^{(p+1/2)\ell(\ell-1)} \frac{1-at^2 q^{2\ell-1}}{(at^2q^{\ell-1};q)_{k+1}} \left[ \begin{array}{c} k \\ \ell \end{array}\right]_q  .
\eea
\end{minipage}
}
\\
\\
The equivalence of two kinds of the expressions are shown in \S\ref{sec:equivalence}. Since this is just a guesswork, at least several checks need to be done.  We find that the following checks in various special cases support the validity of the expressions \eqref{superpoly1}, \eqref{superpoly2}, \eqref{superpoly3} and \eqref{superpoly4}.
\begin{itemize}
\item For $a=q^2$ and $t=-1$, the above formulae reduce to the colored Jones polynomials written in \S\ref{sec:jones}
\item Acting differentials on the formulae, they reproduce known results for homological invariants of the twist knots such as $\fraksl_2$ homological invariants in the fundamental representation and  $s$-invariants \cite{Rasmussen:2004}. $\Longrightarrow$ \S\ref{sec:diff}
\item We checked that  the ``special'' colored superpolynomials which are the limits  $q\to 1$ of the colored superpolynomials obey the  property \cite{Morozov:2012am},
\be\label{special}
\lim_{q\to 1 }\scP_n(K_p;a,q,t) = \left[ \lim_{q\to 1 } \scP_{2}(K_p;a,q,t) \right]^{n-1}. 
\ee
$\Longrightarrow$ \S\ref{sec:sp}
\item For $t=-1$, they reduce to the colored HOMFLY polynomials. We checked they agree with the colored HOMFLY polynomials computed by $SU(N)$ Chern-Simons theory \cite{Zodinmawia:2011ud,Zodinmawia:2012sx} up to 10 crossings. $\Longrightarrow$ \S\ref{sec:10crossings}
\item The colored HOMFLY polynomials can be reformulated into the Ooguri-Vafa polynomials. We checked that the Ooguri-Vafa polynomials obey the conjectural form \cite{Ooguri:1999bv}.  $\Longrightarrow$ \S\ref{sec:OV}

\end{itemize}

\subsection{Differentials}\label{sec:diff}
As briefly explained in \S\ref{sec:intro}, the bi-graded homology such as colored $\fraksl_N$ homology can be realized as the homology of the triply-graded homology with respect to a certain differential of $(a,q,t)$-grading $(\a,\b,\c)$ \cite{Dunfield:2005si,Gukov:2011ry,Fuji:2012pm,Fuji:2012nx}
\bea
\left(\scH_{*,*,*}^{R},d_N\right)\cong \scH_{*,*}^{\fraksl_N,R} \quad  {\rm where} \quad d_N:\scH_{i,j,k}\to \scH_{i+\a,j+\b,k+\c} \ .
\eea
The $(a,q,t)$-grading of the differential $d_N$ ($N>0$ and $N<0$) are as follows.
\be
\begin{array}{c@{\;}|@{\;}c@{\;}|@{\;}c@{\;}c}
\text{differentials} & \text{factors} & (a, q, t)~\text{grading} \\\hline
d_{N>0} & \quad 1 + a^{-1} q^{N} t^{-1} \quad & (-1,N,-1) \\[.1cm]
d_{N<0} & 1 + a^{-1} q^{N} t^{-3} & (-1,N,-3) \\[.1cm]
\end{array}
\label{gradingtabl}
\ee
We should clarify that the differentials $d_{N<0}$ are related to the anti-symmetric representation. 
At the level of Poincar\'e polynomials, the relation amounts to
\be
\scP_n (a,q,t) \; = \;
R^{\fraksl_N}_n (a,q,t) + (1 + a^{-1} q^{N} t^{-1}) Q_n ^{\fraksl_N}(a,q,t) \quad  {\rm for}\ N>0 ,
\label{superright}
\ee
where the ``remainder'' reduces to the Poincar\'e polynomials of the bi-graded homology
\be
R^{\fraksl_N}_n (K;a=q^N,q,t) = \scP^{\fraksl_N}_{n} (K;q,t) \quad {\rm for} \quad  N\ge2 \, .
\ee

\begin{table}
\centering{}
\renewcommand{\arraystretch}{1.5}
\begin{tabular}{|c|p{12cm}|}
\hline 
 ${\bf Knot}$ &  $\scP^{\fraksl_2}_{n=2}(K;q,t)$\tabularnewline
\hline 
\hline
${\bf 4_1}$ &$q^2 t^2+\frac{1}{q^2 t^2}+q t+\frac{1}{q t}+1$
\tabularnewline
\hline
 ${\bf 5_2}$ & $q + q^2 t + 2 q^3 t^2 + q^4 t^3 + q^5 t^4 + q^6 t^5$
 \tabularnewline
\hline
${\bf 6_1}$ & $\frac{1}{q^4 t^4}+\frac{1}{q^3 t^3}+q^2 t^2+\frac{1}{q^2 t^2}+q t+\frac{2}{q t}+2$
\tabularnewline
\hline 
${\bf 7_2}$ & $q + q^2 t + 2 q^3 t^2 + 2 q^4 t^3 + 2 q^5 t^4 + q^6 t^5 + q^7 t^6 + 
 q^8 t^7$
 \tabularnewline
\hline 
${\bf 8_1}$ & $\frac{1}{q^6 t^6}+\frac{1}{q^5 t^5}+\frac{1}{q^4 t^4}+\frac{2}{q^3 t^3}+q^2 t^2+\frac{2}{q^2 t^2}+q t+\frac{2}{q t}+2$
\tabularnewline
\hline 
${\bf 9_2}$ & $q + q^2 t + 2 q^3 t^2 + 2 q^4 t^3 + 2 q^5 t^4 + 2 q^6 t^5 + 
 2 q^7 t^6 + q^8 t^7 + q^9 t^8 + q^{10} t^9$
 \tabularnewline
\hline 
${\bf 10_1}$ & $\frac{1}{q^8 t^8}+\frac{1}{q^7 t^7}+\frac{1}{q^6 t^6}+\frac{2}{q^5 t^5}+\frac{2}{q^4 t^4}+\frac{2}{q^3 t^3}+q^2 t^2+\frac{2}{q^2 t^2}+q t+\frac{2}{q t}+2$
\tabularnewline
\hline 
 \end{tabular}\caption{Poincar\'e polynomials of the $\fraksl_2$ homology in the fundamental representation for the twist knots up to 10 crossings. These are also known as reduced Khovanov polynomials. We should note that after a change of variables $t \rightarrow t^{-1}$ and $q \rightarrow q^{-2}$, our results match with the results obtained by the Mathematica Package {\tt KnotTheory$\grave{}$} \citep{KnotAtlas} up to a multiplicative factor of $q^{-1}$.}
\label{tab:khv}
\end{table}

Since the twist knots $K_p$ are thin knots, we have $Q_n ^{\fraksl_N}(a,q,t)=0$. Therefore, the naive specializations provide the Poincar\'e polynomials of  the colored $\fraksl_N$ homology:
\be
\scP_n (K_p;a=q^N,q,t) = \scP^{\fraksl_N}_{n} (K_p;q,t) \quad {\rm for} \quad N\ge 2 \ . 
\ee 
As a check to see whether the conjectural form \eqref{superpoly1} and \eqref{superpoly2} reproduce the known results for the Poincar\'e polynomials of the $\fraksl_2$ homology $\scP^{\fraksl_2}_{n=2}(K_p;q)$ ({\it a.k.a.} reduced Khovanov polynomials), we have calculated these polynomials  for the twist knots up to 10 crossings and present them in Table \ref{tab:khv}.

Now, let us consider other important differentials called \emph{cancelling differentials} \cite{Dunfield:2005si,Gukov:2011ry,Fuji:2012pm}.\footnote{We thank H. Fuji for suggesting this approach to us.} 
 A canceling differential is characterized in such a way that the homology of the triply-graded theory with respect to this differential is ``trivial", which means  that the consequent homology is one-dimensional in reduced theory.
For instance, it is known that the differential $d_1$ is a canceling differential since the $\fraksl_1$ homology $\scH_{i,j}^{\fraksl_1,R}$ is one-dimensional. (See Figure \ref{fig:chart}.) Besides, the differential $d_{-n}$ which stems from anti-symmetric representation $\Lambda^{n}$ turns out to be another canceling differential.
At this point, we emphasize that the $(a,q,t)$-grading of  the remaining homology with respect to the canceling differential has the defining property:
\bea
\deg\left(\scH^{\scS^{n}}_{*,*,*}(K),d_1\right)&=&\Big(n\,s(K)\,,\,-n\,s(K)\, ,\,0\Big) \ , \cr
\deg\left(\scH^{\scS^{n}}_{*,*,*}(K),d_{-n}\right)&=&\left(n \, s(K)\,,\,n^2\,s(K)\,,\,2n\,s(K)\right) \ , 
\eea
where $s(K)$ is the $s$-invariant of the knot $K$ Rasmussen introduced in \cite{Rasmussen:2004}. 
For the twist knots $K_p$, it is easy to see the action of the differential $d_1$ on the colored superpolynomials
\bea
\scP_{n+1}(K_{p>0};a,q,t)&=&a^{n}q^{-n}+(1+a^{-1}qt^{-1})Q_{n+1}^{\fraksl_1}(K_{p>0};a,q,t) \ ,\cr
\scP_{n+1}(K_{p<0};a,q,t)&=&1+(1+a^{-1}qt^{-1})Q_{n+1}^{\fraksl_1}(K_{p<0};a,q,t)\ ,
\eea
as well as the action of the differential  $d_{-n}$
\bea
\scP_{n+1}(K_{p>0};a,q,t)&=&a^{n}q^{n^2}t^{2n}+(1+a^{-1}q^{-n}t^{-3})Q_{n+1}(K_{p>0};a,q,t) \ ,\cr
\scP_{n+1}(K_{p<0};a,q,t)&=&1+(1+a^{-1}q^{-n}t^{-3})Q_{n+1}(K_{p<0};a,q,t)\ .
\eea
The results are consistent with the $s$-invariants of the twist knots, $s(K_{p>0})=1$ and $s(K_{p<0})=0$.\footnote{We follow the convention of \cite{Gukov:2011ry,Fuji:2012pm,Fuji:2012nx} rather than \cite{Dunfield:2005si,Aganagic:2011sg,Aganagic:2012ne,DuninBarkowski:2011yx}.} Thus, this non-trivial check enforces the justification of the formulae
\eqref{superpoly1} and \eqref{superpoly2} .

\section{Super-$A$-polynomials}\label{sec:super-a-poly}
In this section, we shall find both classical and quantum super-$A$-polynomials of the twist knots $K_p$ for small values of $p$, based on the expression of the colored superpolynomials \eqref{superpoly3} and \eqref{superpoly4}. We will investigate, in the case of the twist knots, the 
conjectures proposed in \cite{Fuji:2012pm,Fuji:2012nx} for colored superpolynomials  which are categorifications of the generalized volume conjecture \cite{Gukov:2003na} and the quantum volume conjecture \cite{Gukov:2003na,Garoufalidis:2003a}. At the end of this section, we shall see the conjecture \cite{Aganagic:2012jb} on the relation between super-$A$-polynomials and augmentation polynomials of knot contact homology.

%For quantum super-$A$-polynomials, we shall use the program developed by Xinyu Sun.
 %In addition, we shall see that the generalized volume conjecture given in \cite{Fuji:2012nx} holds for the twist knots. The critical %points of asymptotic behavior of the colored superpolynomials give the classical super-$A$-polynomials of the twist knots. 

\subsection{Classical super-$A$-polynomials for  twist knots}\label{sec:capoly}

The classical $A$-polynomial $A(K;x,y)$ \cite{Cooper:1994} of a knot $K$ is the character variety of $SL(2,\bC)$-representation $\rho:\pi_1(S^3\backslash K) \to SL(2,\bC)$ of the fundamental group of the knot complement. It encodes rich information about both the topology
and geometry of the knot complement $S^3\backslash K$. The classical $A$-polynomials $A(K_p;x,y)$ of the twist knots $K_p$ were obtained in \cite{Hoste:2004} by finding the recursion relation of the $A$-polynomials $A(K_p;x,y)$ with respect to $p$. (See Table \ref{tab:a poly}.)
\begin{table}
\begin{center}
\begin{tabular}{|c|p{13cm}|}
\hline 
\textbf{Knot}& $A(K;x,y)=A^{\rm super}(K;x,y;a=1,t=-1)$ \tabularnewline
\hline 
\hline \rule{0pt}{5mm}
${\bf 5_2}$ & \footnotesize{$ (y-1)(x^7-x^2(-1+x-2 x^3-2 x^4+x^5) y+(-1+x(2+2 x-x^3+x^4)) y^2+y^3)$}\tabularnewline
\hline\rule{0pt}{5mm}
${\bf 6_1}$&\footnotesize{$(y-1)(1+3 y-2 y x^{-2}+3 y x^{-1}+x^3 y-x^4 y+6 y^2+y^2x^{-4}-3 y^2x^{-3}-y^2x^{-2}+3 y^2x^{-1}+3 x y^2-x^2 y^2-3 x^3 y^2+x^4 y^2+3 y^3-y^3x^{-4}+y^3x^{-3}+3 x y^3-2 x^2 y^3+y^4)$}\tabularnewline
\hline \rule{0pt}{5mm}
${\bf 7_2}$&\footnotesize{$(y-1)(x^{11}+x^4 y-x^5 y+3 x^9 y+4 x^{10} y-2 x^{11} y-2 x^2 y^2+5 x^3 y^2+x^4 y^2-4 x^5 y^2+6 x^7 y^2+5 x^8 y^2+2 x^9 y^2-4 x^{10} y^2+x^{11} y^2+y^3-4 x y^3+2 x^2 y^3+5 x^3 y^3+6 x^4 y^3-4 x^6 y^3+x^7 y^3+5 x^8 y^3-2 x^9 y^3-2 y^4+4 x y^4+3 x^2 y^4-x^6 y^4+x^7 y^4+y^5)$}
\tabularnewline
\hline\rule{0pt}{5mm}
${\bf 8_1}$&\footnotesize{$(y-1)(1+4 y-3 y x^{-2}+5 y x^{-1}+x^5 y-x^6 y+10 y^2+3 y^2x^{-4}-10 y^2x^{-3}+12 y^2x^{-1}+4 x^3 y^2-6 x^5 y^2+2 x^6 y^2+20 y^3-y^3x^{-6}+5 y^3x^{-5}-6 y^3x^{-4}-5 y^3x^{-3}-3 y^3x^{-2}+10 y^3x^{-1}+10 x y^3-3 x^2 y^3-5 x^3 y^3-6 x^4 y^3+5 x^5 y^3-x^6 y^3+10 y^4+2 y^4x^{-6}-6 y^4x^{-5}+4 y^4x^{-3}+12 x y^4-10 x^3 y^4+3 x^4 y^4+4 y^5-y^5x^{-6}+y^5x^{-5}+5 x y^5-3 x^2 y^5+y^6)$}
 \tabularnewline
\hline
 \end{tabular}
\caption{Classical $A$-polynomials for the twist knots. The results were first obtained in \cite{Hoste:2004} where $x=M^2$ and $y=L$. (See also Appendix C in \cite{Garoufalidis:2008a}.) These are equal to the classical super-$A$-polynomials at $a=1$ and $t=-1$.}
\label{tab:a poly}
\end{center}
\end{table}

As in \eqref{volumeconj2}, it turned out \cite{Gukov:2003na} that the classical $A$-polynomial $A(K;x,y)$ is related  to the asymptotic behavior of the colored Jones polynomials $J_n(K,q)$. In \cite{Fuji:2012nx}, the natural generalization of the classical $A$-polynomial $A(K;x,y)$ was proposed for the colored superpolynomial by replacing $J_n(K,q)$ by $\scP_n(K;a,q,t)$ in \eqref{volumeconj2}. This is called the classical super-$A$-polynomial $A^{\rm super}(K;x,y;a,t)$. Thus, let us briefly review how \eqref{volumeconj2} was generalized in \cite{Fuji:2012nx}. 
In the limit
\be
q = e^{\hbar} \to 1 \,, \qquad a = \text{fixed} \,, \qquad t = \text{fixed} \,, \qquad x = q^n = \text{fixed} \ ,
\label{reflimit}
\ee
 the $n$-colored superpolynomials $\scP_n (K;a,q,t)$ exhibit the form in the $n \rightarrow \infty$ limit (also called 
``large color'' behavior):
\be
\scP_n (K;a,q,t) \;\overset{{n \to \infty \atop \hbar \to 0}}{\sim}\;
\exp\left( \frac{1}{\hbar} \int \log y \frac{dx}{x} \,+\, \ldots \right)
%\exp\left( \frac{1}{\hbar} S_0 (x,a,t) \,+\,\sum_{n=0}^\infty S_{n+1} (x,a,t) \, \hbar^{n} \right)
\label{VCsuper}
\ee
where the ellipsis denote subleading terms  which are regular in the $\hbar \to 0$ limit. 
The leading term is given by the integral on the zero locus of the classical super-$A$-polynomial,
\be
A^{\text{super}} (x,y;a,t) \; = \; 0 \,.
\label{supercurve}
\ee
Following this procedure performed in \cite{Fuji:2012nx}, we
shall obtain the classical super-$A$-polynomials $A^{\rm super}(K_p;x,y;a,t)$ of the twist knots $K_p$ for small values of $p$.

To demonstrate the explicit calculation, we introduce the two  variables  $z=e^{\hbar k}$ and 
$w=e^{\hbar \ell}$. Then, in the limit \eqref{reflimit},
the sum over $k$ and $\ell$ in \eqref{superpoly3} can be approximated by the integral over $z$ and $w$:
\be
\scP_{n}(K_p;a,q,t) \; \sim \; \int dzdw\; e^{\frac{1}{\hbar}\left(\widetilde\cW(K_p;z,w)+{\cal O}(\hbar)\right)} \,, \phantom{\int^1}
\label{Pn-saddle}
\ee
with the potential function $\widetilde\cW(z,w)$ for $K_{p>0}$
\begin{eqnarray}\label{potential}
 \widetilde\cW(K_{p>0};z,w)
&=&  -\log x\log (-t)- \frac{\pi^2}{3} +i\pi \log w+p(\log a  + 2 \log t) \log w+ \left(p+\tfrac12\right) (\log w)^2 \cr
& &    + \Li_2( x^{-1}) - \Li_2(x^{-1}z) + \Li_2(-a t) - \Li_2(-a t z) + \Li_2(-a t^3 x ) \cr
&&-  \Li_2(-a t^3 xz)- \Li_2(a t^2 w)+ \Li_2(a t^2 wz)+\Li_2(w)+\Li_2(zw^{-1}) \ ,
\end{eqnarray}
and, for $K_{p<0}$, we just drop the first term of \eqref{potential}
\begin{eqnarray}\label{potential2}
 \widetilde\cW(K_{p<0};z,w)
&=& - \frac{\pi^2}{3} +i\pi \log w+p(\log a  + 2 \log t) \log w+ \left(p+\tfrac12\right) (\log w)^2 \cr
& &    + \Li_2( x^{-1}) - \Li_2(x^{-1}z) + \Li_2(-a t) - \Li_2(-a t z) + \Li_2(-a t^3 x ) \cr
&&-  \Li_2(-a t^3 xz)- \Li_2(a t^2 w)+ \Li_2(a t^2 wz)+\Li_2(w)+\Li_2(zw^{-1}) \, .
\end{eqnarray}
These can be regarded as the refined versions of the Neumann-Zagier potentials \cite{Neumann:1985}, which can be obtained by applying the saddle point approximation \cite{Hikami:2007zz}
to \eqref{superpoly3}
and by using the asymptotics of the $q$-Pochhammer symbol
\begin{eqnarray}
(z;q)_{k}\sim e^{\frac{1}{\hbar}\left({\rm Li}_2(z)-{\rm Li}_2(zq^k)\right)} \,.
\end{eqnarray}
The leading asymptotic behavior \eqref{VCsuper} with respect to $\hbar$ comes from the saddle point
\bea
\frac{\partial \widetilde{\mathcal{W}}(K_p;z,w,x)}{\partial z}\Bigg|_{(z,w)=(z_0,w_0)}=0=\frac{\partial \widetilde{\mathcal{W}}(K_p;z,w,x)}{\partial w}\Bigg|_{(z,w)=(z_0,w_0)}\,.
\eea
Hence, the zero locus~\eqref{supercurve}  of the classical super-$A$-polynomial are determined by
\be
y = \exp\left(x\frac{\partial \widetilde{\mathcal{W}}(K_p;z_0,w_0,x)}{\partial x}\right) \,.    \label{yV}
\ee
Plugging the expression \eqref{potential} to the above two equations, we obtain the following system
\bea
&&1 = \frac{w_0(x - z_0) (1 + a t z_0) (1 + a t^3 x z_0)}{x (w_0 - z_0)(1 - a t^2  w_0z_0)}  \ ,  \\
&&1 =- \frac{a^p t^{2p} w_0^{2p} (w_0-z_0)(1 - a t^2  w_0)}{ (1-w_0) (1 - a t^2  w_0z_0)} \ ,   \\
&&y = - \frac{(x-1)(1 + a t^3 x z_0) }{t(x - z_0)(1 + a t^3 x) }  \ , \label{lasteq}
\eea
for $K_{p>0}$, and the last equation \eqref{lasteq} is replaced by
\bea
y = \frac{(x-1)(1 + a t^3 x z_0) }{(x - z_0)(1 + a t^3 x) }  
\eea
for $K_{p<0}$.
Eliminating $z_0, w_0$ from these three equations, we indeed reproduce the classical super-$A$-polynomial.\footnote{We should note that we use the command {\tt eliminate} in Maple for this operation. We thank H. Fuji for explaining us.} The results are summarized in Table \ref{tab:cl super a poly 1}, Table \ref{tab:cl super a poly 2} and Table \ref{tab:cl super a poly 3}. Then, it is straightforward to see that, at $a=1$ and $t=-1$, the classical super-$A$-polynomials reduce to the classical $A$-polynomials obtained in \cite{Hoste:2004}. (See Table \ref{tab:a poly}.)

\begin{table}
\begin{center}
\begin{tabular}{|c|p{13.5cm}|}
\hline 
\textbf{Knot}& $A^{\rm super}(K;x,y;a,t)$ \tabularnewline
\hline 
\hline 
${\bf 5_2}$ & \footnotesize{$y^4$}\\[3pt] \rule{0pt}{2mm}
&$-\frac{a}{1+a t^3 x}$\footnotesize{$(2-x+t x-2 t^2 x+3 t^2 x^2+a t^2 x^2+4 a t^3 x^2-2 a t^3x^3+2 a t^4 x^3+2 a t^5 x^3-a t^5 x^4+2 a^2 t^5 x^4+2 a^2 t^6 x^4-a^2 t^6 x^5+a^2 t^7 x^5+a^3 t^8 x^6) y^3$}\\[3pt] \rule{0pt}{2mm}
&$-\frac{a^2 (-1+x)}{\left(1+a t^3 x\right)^2 y^4}$\footnotesize{$(1+t x-2 t^2 x+2 t^2 x^2-2 t^3 x^2+4 a t^3 x^2+t^4 x^2-3 t^4 x^3+a t^4 x^3-2 a t^5 x^3+4 a t^5 x^4-4 a t^6 x^4+6 a^2 t^6 x^4-4 a t^7 x^4+3 a t^7 x^5-a^2 t^7 x^5+2 a^2 t^8 x^5+2 a^2 t^8 x^6-2 a^2 t^9 x^6+4 a^3 t^9 x^6+a^2 t^{10} x^6-a^3 t^{10} x^7+2 a^3 t^{11} x^7+a^4 t^{12} x^8) y^2$}\\[3pt] \rule{0pt}{2mm}
&$+\frac{a^3 t^3 (-1+x)^2 x^2}{\left(1+a t^3 x\right)^3}$\footnotesize{$(1+t x-t^2 x-t^3 x^2+2 a t^3 x^2+2 a t^4 x^2+2 a t^4 x^3-2 a t^5 x^3-2 a t^6 x^3+3 a t^6 x^4+a^2 t^6 x^4+4 a^2 t^7 x^4+a^2 t^7 x^5-a^2 t^8 x^5+2 a^2 t^9 x^5+2 a^3 t^{10} x^6) y$}\\[3pt] \rule{0pt}{2mm}
&$-\frac{a^5 t^{11} (-1+x)^3 x^7}{\left(1+a t^3 x\right)^3}$
\tabularnewline
\hline
 \end{tabular}
\caption{Classical super-$A$-polynomial of the knot ${\bf 5_2}$}
\label{tab:cl super a poly 1}
\end{center}
\end{table}

From these tables of the classical super-$A$-polynomials, we observe, at $x=1$, the following relation between the super-$A$-polynomials and the superpolynomials
\be\label{strange id}
 A^{\rm super}(K_p;x=1,y;a,t) = y^k\left(y-\scP(K_p;a,q=1,t)\right) ~.
\ee
The same relation up to relative sign was found in \cite{Fuji:2012nx} for the torus knots $T^{2,2p+1}$. (See (1.19) in \cite{Fuji:2012nx}.) At $t=-1$, the super-$A$-polynomials reduce to the $Q$-deformed $A$-polynomials. (See Table \ref{tab:notation}.) At this level, the identity \eqref{strange id} relates $Q$-deformed $A$-polynomial to the HOMFLY polynomials:
\be
A^{Q}(K_p;x=1,y;a) = y^k\left(y-P(K_p;a,q=1)\right)~.
\ee
This relation may have something to do with the large-$N$ dulaity \cite{Aganagic:2012jb} which needs further investigation.

\subsection{Quantum super-$A$-polynomials for twist knots}

For the twist knots, the AJ conjecture have also been studied since the pioneering paper \cite{Garoufalidis:2003a,Garoufalidis:2003b}. As explained in \S\ref{sec:intro}, the quantum $A$-polynomials $\widehat A(K;\hat x,\hat y;q)$ are obtained by the difference equation of the colored Jones polynomials $J(k;q)$ of minimal order.
For a knot whose colored Jones  polynomial  involves a single summation, the difference equation 
can be determined \cite{Garoufalidis:2003a} by using the Mathematica package, {\tt qZeil.m} \cite{qZeil}.  We have seen that the colored
Jones polynomials of the twist knots $K_p$ involves  the double summation \eqref{jones twist 3}. Therefore, the package {\tt qZeil.m} cannot be used  for  the twist knots  $K_p$ where  $|p|>1$.\footnote{Even for a multi-sum fomula, a difference equations can be obtained \cite{Takata:2004} by using  the Mathematica package {\tt qMultisum.m} \cite{qZeil}. However, as pointed out in \cite{Garoufalidis:2008a}, 
 the package {\tt qMultisum.m} does not give a difference equation of minimal  order  in general.}  The algorithm which computes difference equations of minimal order for multi-sum expressions has been developed  by S. Garoufalidis and X. Sun \cite{Garoufalidis:2005,Garoufalidis:2008a,Garoufalidis:2008b}. 
By the
 algorithm with a certi�cate,  the quantum $A$-polynomials  were obtained for the twist knots  $\widehat{A}(K_p;\hat x,\hat y; q)$  in the
range $-14\leq p \leq 15$ \cite{Garoufalidis:2008a}.

\begin{table}
\centering{}
\begin{tabular}{|c|p{13.5cm}|}
\hline 
{\small ${\bf Knot}$} &  $\widehat{A}^{{\rm super}}(K;\hat{x},\hat{y};a,q,t)$\tabularnewline
\hline 
\hline 
${\bf 5_{2}}$
&\footnotesize{$q^{-31}(q^2+a t^3 \hat{x})(q^3+a t^3 \hat{x})(q^4+a t^3 \hat{x})(q^7+a t^3 \hat{x}^2)(q^8+a t^3 \hat{x}^2)(q^9+a t^3 \hat{x}^2) \hat{y}^4$}
\\[5pt] \rule{0pt}{2mm}
&\footnotesize$-a{q^{-27}}(q^2+a t^3 \hat{x})(q^3+a t^3 \hat{x})(q^2+a t^3 \hat{x}^2)(q^6+a t^3 \hat{x}^2)(q^7+a t^3 \hat{x}^2)(q^8+q^9-q^8 \hat{x}+q^8 t \hat{x}-q^7 t^2 \hat{x}-q^8 t^2 \hat{x}+q^6 t^2 \hat{x}^2+a q^6 t^2 \hat{x}^2+q^7 t^2 \hat{x}^2+q^8 t^2 \hat{x}^2+a q^3 t^3 \hat{x}^2+a q^4 t^3 \hat{x}^2+a q^7 t^3 \hat{x}^2+a q^8 t^3 \hat{x}^2-a q^3 t^3 \hat{x}^3-a q^7 t^3 \hat{x}^3+a q^3 t^4 \hat{x}^3+a q^7 t^4 \hat{x}^3+a q^4 t^5 \hat{x}^3+a q^5 t^5 \hat{x}^3+a^2 q t^5 \hat{x}^4-a q^4 t^5 \hat{x}^4+a^2 q^5 t^5 \hat{x}^4+a^2 q^2 t^6 \hat{x}^4+a^2 q^3 t^6 \hat{x}^4-a^2 q^2 t^6 \hat{x}^5+a^2 q^2 t^7 \hat{x}^5+a^3 t^8 \hat{x}^6) \hat{y}^3$
\\[5pt] \rule{0pt}{2mm}
&\footnotesize$-a^2{q^{-16}}(-1+\hat{x})(q^2+a t^3 \hat{x})(q^2+a t^3 \hat{x}^2)(q^5+a t^3 \hat{x}^2)(1+a q t^3 \hat{x}^2)(q^8+q^8 t \hat{x}-q^7 t^2 \hat{x}-q^8 t^2 \hat{x}+q^7 t^2 \hat{x}^2+q^8 t^2 \hat{x}^2+a q^4 t^3 \hat{x}^2+a q^5 t^3 \hat{x}^2-q^7 t^3 \hat{x}^2+a q^7 t^3 \hat{x}^2-q^8 t^3 \hat{x}^2+a q^8 t^3 \hat{x}^2+q^7 t^4 \hat{x}^2+a q^4 t^4 \hat{x}^3-q^6 t^4 \hat{x}^3-a q^6 t^4 \hat{x}^3-q^7 t^4 \hat{x}^3-q^8 t^4 \hat{x}^3+a q^8 t^4 \hat{x}^3-a q^3 t^5 \hat{x}^3-a q^4 t^5 \hat{x}^3+a q^5 t^5 \hat{x}^3+a q^6 t^5 \hat{x}^3-a q^7 t^5 \hat{x}^3-a q^8 t^5 \hat{x}^3+a q^3 t^5 \hat{x}^4+a q^4 t^5 \hat{x}^4+a q^7 t^5 \hat{x}^4+a q^8 t^5 \hat{x}^4+a^2 q t^6 \hat{x}^4-a q^3 t^6 \hat{x}^4+a^2 q^3 t^6 \hat{x}^4-a q^4 t^6 \hat{x}^4+2 a^2 q^4 t^6 \hat{x}^4+a^2 q^5 t^6 \hat{x}^4-a q^7 t^6 \hat{x}^4+a^2 q^7 t^6 \hat{x}^4-a q^8 t^6 \hat{x}^4-a q^4 t^7 \hat{x}^4-2 a q^5 t^7 \hat{x}^4-a q^6 t^7 \hat{x}^4-a^2 q^2 t^7 \hat{x}^5+a q^4 t^7 \hat{x}^5+a^2 q^4 t^7 \hat{x}^5+a q^5 t^7 \hat{x}^5+a q^6 t^7 \hat{x}^5-a^2 q^6 t^7 \hat{x}^5+a^2 q t^8 \hat{x}^5+a^2 q^2 t^8 \hat{x}^5-a^2 q^3 t^8 \hat{x}^5-a^2 q^4 t^8 \hat{x}^5+a^2 q^5 t^8 \hat{x}^5+a^2 q^6 t^8 \hat{x}^5+a^2 q^3 t^8 \hat{x}^6+a^2 q^4 t^8 \hat{x}^6+a^3 t^9 \hat{x}^6+a^3 q t^9 \hat{x}^6-a^2 q^3 t^9 \hat{x}^6+a^3 q^3 t^9 \hat{x}^6-a^2 q^4 t^9 \hat{x}^6+a^3 q^4 t^9 \hat{x}^6+a^2 q^3 t^{10} \hat{x}^6-a^3 q^2 t^{10} \hat{x}^7+a^3 q t^{11} \hat{x}^7+a^3 q^2 t^{11} \hat{x}^7+a^4 t^{12} \hat{x}^8) \hat{y}^2$
  \\[5pt] \rule{0pt}{2mm}
  &\footnotesize$+a^3 {q^{-6}}t^3(-1+\hat{x}) \hat{x}^2(-1+q \hat{x})(q^2+a t^3 \hat{x}^2)(1+a q^2 t^3 \hat{x}^2)(1+a q^3 t^3 \hat{x}^2) (q^5+q^5 t \hat{x}-q^5 t^2 \hat{x}+a q^2 t^3 \hat{x}^2-q^5 t^3 \hat{x}^2+a q^6 t^3 \hat{x}^2+a q^3 t^4 \hat{x}^2+a q^4 t^4 \hat{x}^2+a q^2 t^4 \hat{x}^3+a q^6 t^4 \hat{x}^3-a q^2 t^5 \hat{x}^3-a q^6 t^5 \hat{x}^3-a q^3 t^6 \hat{x}^3-a q^4 t^6 \hat{x}^3+a q^3 t^6 \hat{x}^4+a^2 q^3 t^6 \hat{x}^4+a q^4 t^6 \hat{x}^4+a q^5 t^6 \hat{x}^4+a^2 t^7 \hat{x}^4+a^2 q t^7 \hat{x}^4+a^2 q^4 t^7 \hat{x}^4+a^2 q^5 t^7 \hat{x}^4+a^2 q^3 t^7 \hat{x}^5-a^2 q^3 t^8 \hat{x}^5+a^2 q^2 t^9 \hat{x}^5+a^2 q^3 t^9 \hat{x}^5+a^3 q t^{10} \hat{x}^6+a^3 q^2 t^{10} \hat{x}^6) \hat{y}$
  \\[5pt] \rule{0pt}{2mm}
  &\footnotesize$-a^5 q^3 t^{11} (-1+\hat{x}) \hat{x}^7 (-1+q \hat{x})(-1+q^2 \hat{x}) (1+a q^3 t^3 \hat{x}^2)(1+a q^4 t^3 \hat{x}^2) (1+a q^5 t^3 \hat{x}^2)$
  \tabularnewline
\hline 
\end{tabular}\caption{Quantum super-$A$-polynomial of the knot ${\bf 5_{2}}$.}
\label{tab:qapoly52}
\end{table}

Similarly to the generalized volume conjecture in \S\ref{sec:capoly}, the AJ conjecture was generalized to the following form \cite{Fuji:2012nx} for colored superpolynomials.
As in \eqref{AJ}, the quantum version $\widehat A^{\text {super}}(\hat x,\hat y,q,a,t)$ of  super-$A$-polynomial dependent on
two commutative parameters $(a,t)$ and a quantum parameter $q$  annihilate colored superpolynomial $\scP_n(K;a,q,t)$:
\be
\widehat A^{\text{super}}(K;\hat x, \hat y; a,q,t) \; \scP_n (K;a,q,t) \; = \; 0 \,.   \label{AsuperP}
\ee
%which is a quantum version of the classical curve \eqref{supercurve}.
The two operators $\hat x$ and $\hat y$ act on $\scP_n (K;a,q,t)$ as in \eqref{xyactionJ}:
\bea
\hat x \,  \scP_n (K;a,q,t) =q^n \scP_n (K;a,q,t) \, , \quad \hat y\, \scP_n (K;a,q,t)= \scP_{n+1} (K;a,q,t) \ .
\eea
If we could  obtain a difference equation of minimal order
\be
b_k \, \scP_{n+k} (K;a,q,t) + \ldots + b_1 \, \scP_{n+1} (K;a,q,t) + b_0 \, \scP_n (K;a,q,t) \; = \; 0
\label{QVCsuper}
\ee
where $b_i \equiv b_i (\hat x, a, q, t)$ are  rational functions, then we could read off   
the quantum (non-commutative) super-$A$-polynomial as follows:
\be
\widehat A^{\text{super}} (K;\hat x, \hat y; a,q,t) \; = \; \sum_i b_i (\hat x, a, q, t) \, \hat y^i \,.
\label{Asuperform}
\ee
Taking the classical limit $q \rightarrow 1$, it is conjectured to reduce to the classical super-$A$-polynomial  $A^{\text{super}} (K;x,y;a,t)$ in \eqref{supercurve}.

To find the difference equations of minimal order for the colored superpolynomials of the twist knots, we use the program {\tt iSumq.txt}  written by Xinyu Sun, based on $q$-analogue of the algorithm developed in \cite{Garoufalidis:2008b}. We present the results for the twist knots $K_2={\bf 5_2}$ and $K_{-2}={\bf 6_1}$ in  Table \ref{tab:qapoly52}  and Table \ref{tab:qapoly61} respectively.  Besides, the quantum super-$A$-polynomials for $K_3={\bf 7_2}$ and $K_{-3}={\bf 8_1}$ are available in the ancillary file. In fact, the classical limit $q \to 1$ of the quantum super-$A$-polynomials for these knots agree with 
the classical  super-$A$-polynomials in the previous subsection \S\ref{sec:capoly}.
Furthermore, we have varified that, for these knots,  
the  quantum super-$A$-polynomials  $\widehat A^{\text {super}}(\hat x,\hat y,q,a,t)$  reduce to the quantum $A$-polynomials $\widehat A(\hat x,\hat y;q)$ obtained in \cite{Garoufalidis:2008a}.

\subsection{Quantizability}

In this subsection, we discuss quantizability of the classical super-$A$-polynomials. For more detail, we refer to the reader to \cite{Fuji:2012pm,Fuji:2012nx}.
In order for a classical super-$A$-polynomial $A^{\rm super}(K;x,y;a,t)$ to be quantizable, the integral of the one-form  $ \log y \frac{dx}{x}$  along a path $\gamma$ on the algebraic curve $A^{\rm super}(K;x,y;a,t)=0$ is required to obey the Bohr-Sommerfeld condition \cite{Gukov:2003na,Gukov:2011qp}.
This condition amounts to the following constraints on the periods of the imaginary and real parts of $\log y \frac{dx}{x}$, respectively,
\bea
\oint_{\gamma} \Big( \log |x| d ({\rm arg} \, y) - \log |y| d ({\rm arg} \, x) \Big) & = & 0 \,, \cr
\frac{1}{4 \pi^2} \oint_{\gamma} \Big( \log |x| d \log |y| + ({\rm arg} \, y) d ({\rm arg} \, x) \Big) & \in & \mathbb{Q} \, .
\label{qcondQ}
\eea
for any closed path $\gamma$ on $A^{\rm super}(K;x,y;a,t)=0$. By interpreting these constraints in the context of algebraic K-theory \cite{Gukov:2011qp}, it turns out that the classical super-$A$-polynomial $A^{\rm super}(K;x,y;a,t)$ is quantizable \emph{only if} the classical super-$A$-polynomial $A^{\rm super}(x,y;a,t)$ is tempered. By definition, a polynomial $A(x,y)$ is tempered if all roots of all face polynomials of its Newton polygon are roots of unity.

\begin{figure}[ht]
  \begin{minipage}[b]{8cm}\centering
\includegraphics[scale=0.4]{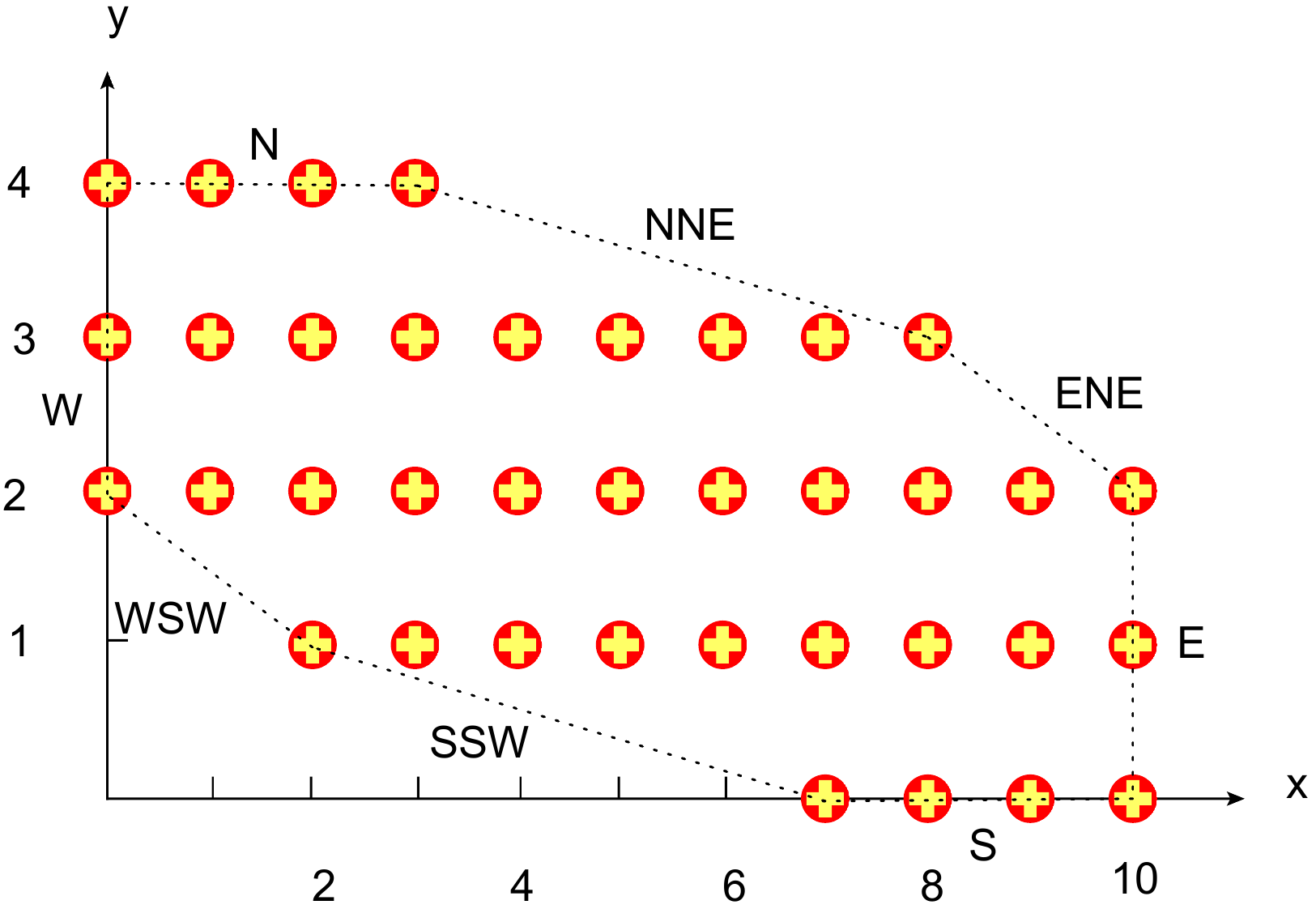}\caption{Newton polygon of super-$A$-polynomial for the knot ${\bf 5_{2}}$. }
\label{fig:newton52}
\end{minipage}
\hspace{.5cm}
  \begin{minipage}[b]{8cm}\centering
\includegraphics[scale=0.4]{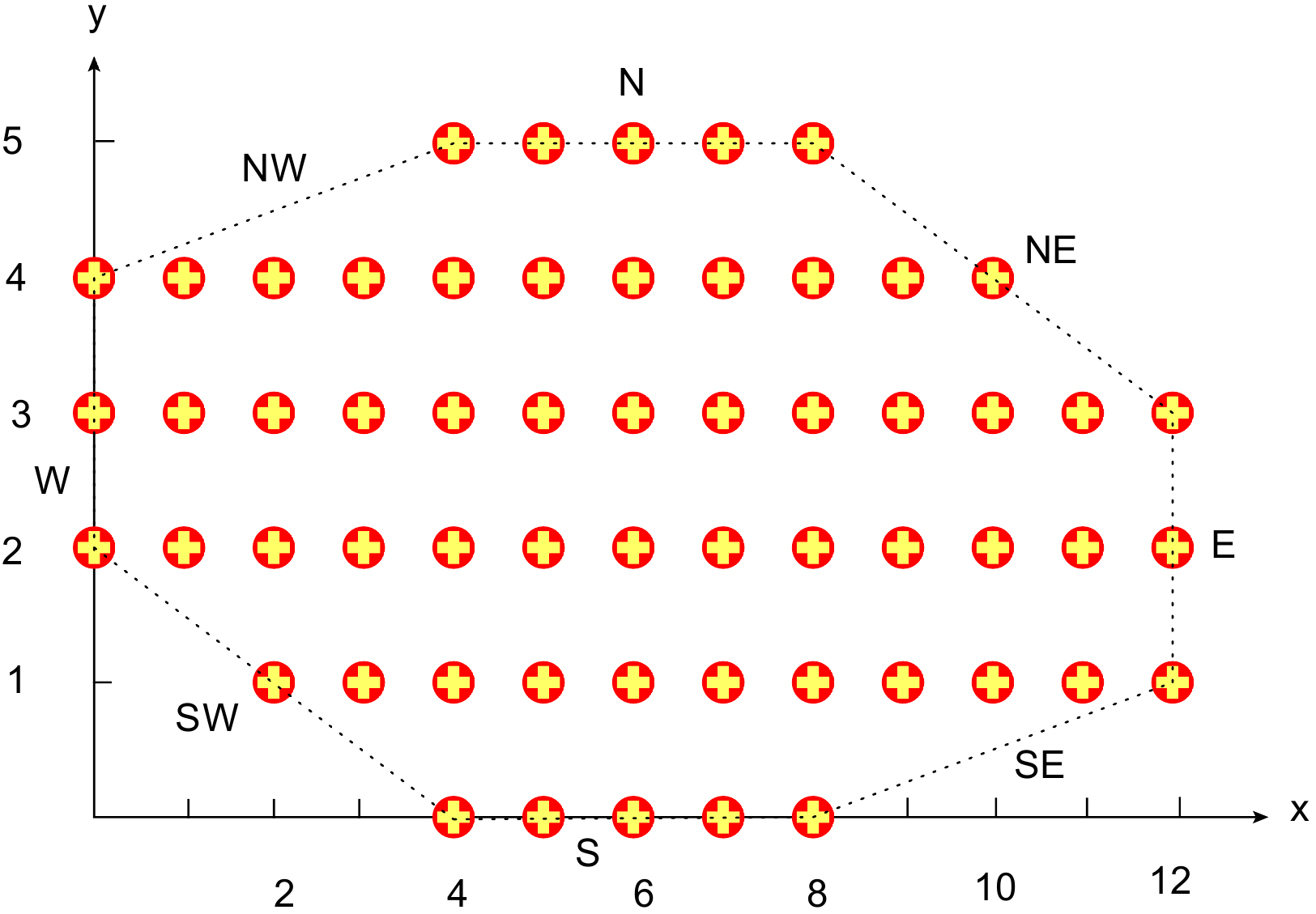}\caption{Newton polygon of super-$A$-polynomial for the knot ${\bf 6_{1}}$. }
\label{fig:newton61}
\end{minipage}
\end{figure}

The Newton polygon of the super-$A$-polynomial $A^{\rm super}(K;x,y;a,t)$ and its face polynomials are constructed as follows. First, we write the matrix $\{c_{ij}\}_{i,j}$ for the super-$A$-polynomial $A^{\rm super}(K;x,y;a,t)= \sum_{i,j} c_{i,j} (a,t) x^i y^j $. (See Figure \ref{fig:matrixform}.\footnote{For the twist knots $K_{p=\pm3}$, the matrix forms of the super-$A$-polynomials are  written in the ancillary notebook.}) Then, the Newton polygon is designed by plotting red circles for monomials $c_{ij}\neq0$ and yellow crosses for monomials at the special limit $c_{ij}(a=1,t=-1)\neq0$. (See Figure \ref{fig:newton52} and Figure \ref{fig:newton61}.) The faces of the Newton polygons are denoted by the dotted line in Figure \ref{fig:newton52} and Figure \ref{fig:newton61}. For a given face, we rename the monomial coefficients on the face as $c_k$. Then, the face polynomial is defined to be  $f(z)=\sum_k c_k z^k$. Assuming that the variables $a,t$ are roots of unity, the quantizability condition requires that all roots of $f(z)$ constructed
for all faces of the Newton polygon must be roots of unity.

Even though the super-$A$-polynomials for the twist knots $K_p$ are not obtained for all $p$, the pattern of the face polynomials can been seen from the matrix forms of the super-$A$-polynomials $A^{\rm super}(K_{p};x,y;a,t)$ for $|p|\le3$, which are summarized in Table \ref{tab:face pos} and Table \ref{tab:face neg}. Therefore, it is easy to see that the classical super-$A$-polynomials $A^{\rm super}(K_{p};x,y;a,t)$  of the twist knots $K_p$ for any $p$ satisfy the necessary condition of quantizability since we assume that  the variables $a,t$ are roots of unity.

\begin{table}[ht]
\begin{minipage}[b]{8cm}\centering
\begin{tabular}{|c|p{5cm}|}
\hline 
 $\text{\bf Face}$ &  $\text{\bf Face polynomials}$\tabularnewline
\hline 
\hline
 N & $-(z+at^3)^{2p-1}$
 \tabularnewline
\hline
NNE & $-a^{2p-1}t^{3(2p-1)}(z-a^{p+1}t^{2p+1})$
\tabularnewline
\hline 
ENE & $a^{3p}t^{8p-2}(z+at)^{p-1}$
 \tabularnewline
\hline 
E & $-a^{3p-1}t^{7p-3}(z-at^2)^{p}$
\tabularnewline
\hline 
S & $(z-1)^{2p-1}$
 \tabularnewline
\hline 
SSW & $(-1)^{p-1}a^{2p-1}t^{3(p-1)}(z+a^pt^{4p})$
\tabularnewline
\hline 
WSW & $(-1)^{p-1}a^{p}(z+a t^3)^p$
\tabularnewline
\hline 
W &$-(z-a)^p$
\tabularnewline
\hline
 \end{tabular}\caption{Face polynomials of $K_{p>0}$}
\label{tab:face pos}
\end{minipage}
\hspace{0.5cm}
\begin{minipage}[b]{8cm}
\centering
\begin{tabular}{|c|p{4cm}|}
\hline 
 $\text{\bf Face}$ &  $\text{\bf Face polynomials}$\tabularnewline
\hline 
\hline
 N & $-(z+at)^{p}$
 \tabularnewline
\hline
NE & $-z+a^pt^{2p}$
\tabularnewline
\hline 
E & $a^pt^{2p}(z+at^3)^{2p}$
 \tabularnewline
\hline 
SE & $a^{3p}t^{8p}(z-at^2)^{p}$
\tabularnewline
\hline 
S & $(-1)^{p}a^{2p}t^{9p}(z+at)^{p}$
 \tabularnewline
\hline 
SW & $(-1)^{p-1}a^{2p}t^{5p}(z-a^pt^{4p})$
\tabularnewline
\hline 
W & $-a^{p}t^{p}(z-1)^{2p}$
\tabularnewline
\hline 
NW &$-(z-at^4)^p$
\tabularnewline
\hline
 \end{tabular}\caption{Face polynomials of $K_{p<0}$}
 \label{tab:face neg}
\end{minipage}
\end{table}

\subsection{Augmentation polynomials of knot contact homology}
In this subsection, we find the relation between $Q$-deformed $A$-polynomials and  augmentation polynomials of knot contact homology in the case of the twist knots. Recently, it was proposed in \cite{Aganagic:2012jb} that the $Q$-deformed $A$-polynomials can be identified to the augmentation polynomial of knot contact homology by change of variables. The explanation of the augmentation polynomial of knot contact homology is beyond the ability of the authors so that we refer to the reader to \cite{Ng:2010,Ekholm:2010,Fuji:2012nx}. Instead, we make an observation on this conjecture for the twist knots.

To relate $Q$-deformed $A$-polynomial to augmentation polynomial for the torus knots $T^{2,2p+1}$, the  identification
\be 
x=-\mu,\,\,y=\frac{1+\mu}{1+U\mu}\lambda,\,\,a=U,\,\,t=-1,\,\,V=1
\ee
was used in  \citep{Fuji:2012nx}. The same identification for the twist knots leads to the results tabulated in Table \ref{tab:augmented}. 
By extrapolating the results for $|p|\le3$, we find the following relation:
\bea
&&A^{\rm{super}}\left(K_{p>0};x=-\mu,y=\frac{1+\mu}{1+U\mu}\lambda;a=U,t=-1\right)\cr&&\hspace{3cm}=\frac{(1+\mu)^{2p-1}}{(1+U\mu)^{2p}}{\rm Aug}(K_{p>0};\mu,\lambda;U,V=1) \ ,\cr
&&A^{\rm{super}}\left(K_{p<0};x=-\mu,y=\frac{1+\mu}{1+U\mu}\lambda;a=U,t=-1\right)\cr&&\hspace{3cm}=\frac{(1+\mu)^{2|p|}}{U^{|p|}(-\mu )^{2|p|}(1+U\mu)^{2|p|+1}}{\rm Aug}(K_{p<0};\mu,\lambda;U,V=1) \ ,
\eea
where we use the results in \cite{Ng:2012} for the augmentation polynomials ${\rm Aug}(K;\mu,\lambda;U,V=1)$.

\section{Discussions}\label{sec:discussions}
In this paper,  we conjecture the forms of the colored superpolynomials  of the twist knots $K_p$ by guessing the twisting factors from the available data. These become the natural generalizations of the colored Jones polynomials. We check in \S\ref{sec:diff} that, by applying the differentials, the formulae reproduce the reduced Khovanov polynomials and the $s$-invariants. This non-trivial check enforces the fact that the refinement parameter $t$ is properly encoded in the expressions we propose. Furthermore, the checks carried out in \S\ref{sec:10crossings} and \S\ref{sec:OV} at the level of the colored HOMFLY polynomials indicate that the conjecture incorporates the parameter $a$ in the right manner. 

Using the program written by Xinyu Sun, we have found the difference equations which result in the quantum super-A-polynomials of the knots $\bf 5_2$ and $\bf 6_1$. 
The results support the categorified versions of the generalized volume conjecture and  the quantum volume conjecture for the colored superpolynomials proposed in \cite{Fuji:2012nx,Fuji:2012pm} since the quantum super-$A$-polynomials in the classical limit  $q \rightarrow 1$ agree with the classical super-$A$-polynomials obtained by the asymptotic behaviors of the colored superpolynomials. Moreover, the analysis of face polynomials ensure that the classical super-$A$-polynomials satisfy the necessary condition of quantizability. In addition, we observed that the conjecture recently proposed by Aganagic and Vafa \cite{Aganagic:2012jb} still holds for the twist knots.

The direct line to study further is to find the expressions of the colored superpolynomials of the twist knots with other representations. Especially, the triply-graded homology with the symmetric representation is related to  that with the anti-symmetric representation via mirror symmetry \cite{Gukov:2011ry}. Also, it is desirable to find the colored superpolynomials of other knots and links as well as  with gauge groups of different type.

This study also provides the following insight for future research. Since the twist knots $K_{p\neq1}$ are all hyperbolic knots, the explicit calculations \eqref{CS inv} of the colored HOMFLY polynomials involve quantum $SU(N)$ Racah coefficients. Therefore, in principle, one can extract the information about the quantum $SU(N)$ Racah coefficients \cite{Nawata:2012} from the conjectured colored superpolynomial at $t=-1$. This would provide an important clue to the long-standing problem about quantum $6j$-symbols for $U_q(\fraksl_N)$.
In fact, a closed form expression for quantum $SU(N)$ Racah coefficients  would enable us to  evaluate $V_R^{SU(N)}$ for any knot. 

It is well known \cite{Lickorish:1962} that every closed oriented three-manifold $M$ can be constructed by Dehn surgery of an framed link $L$ in $S^3$. Based on the data of the Dehn surgery, the Chern-Simons invariant of $M$ known as the WRT invariant \cite{Reshetikhin:1991tc,Kirby:1989zc,Kaul:2000xe} is essentially written as a sum over colors of link invariants  with the modular $S$-matrices. However, it requires further study to obtain the refined Chern-Simons invariants of three-manifolds explicitly.  Especially, it would be interesting to see if the refined Chern-Simons invariants of integral homology three-spheres can be interpreted from the view point of (mock) modular forms \cite{Lawrence:1999}.  We hope to report on this issue in future.

\subsubsection*{Note added}
We were informed of related work done by
H. Fuji, S. Gukov, M. Sto$\check{\rm s}$i$\grave{\rm c}$ and P. Su{\l}kowski \cite{Fuji:2012sx} so that we coordinated the submision
of our paper on arXiv with theirs.

\subsubsection*{Note added 2}
We would be happy to have constructive comments, suggestions, criticisms and reference requests especially from mathematicians.

\section*{Acknowledgement}
The authors would like to thank Tudor Dimofte, Davide Gaiotto, Sergei Gukov, Alexei Morozov, Lenny Ng, Pavel Putrov, Marko Sto$\check{\rm s}$i$\grave{\rm c}$, Piotr Su{\l}kowski, and especially Hiroyuki Fuji for useful correspondence and discussions. We are really grateful to  Stavros Garoufalidis for the correspondence on the program for quantum $A$-polynomials. 
 This research was initiated during  ``$N=2$ JAZZ workshop'' at McGill University. S.N. and P.R. would like to thank the organizers of the workshop for its warm hospitality. S.N. is also indebted to ``2012 Summer Simons Workshop in Mathematics and Physics'' at Simons Center and ``New Perspectives in Topological Field Theories'' sponsored by ITGP and ESF RNP at Center for Mathematical Physics, Hamburg for providing stimulating research environments. The work of S.N. is supported by the ERA Grant ER08-05-174 and  by the European Union Seventh Framework Programme FP7-People-2010-IRSES under grant agreement n$^\circ 269217$. Research at Perimeter Institute is supported by the Government of Canada through Industry Canada and by the Province of Ontario through the Ministry of Research and Innovation.

\appendix

\section{Equivalence of two expressions}\label{sec:equivalence}
In this appendix, we shall show the multi-sum expression \eqref{superpoly1} is equal to the double-sum expression \eqref{superpoly3}. The proof is based on the technique called  \emph{Bailey chains} \cite{Andrew:1984}. In what follows,  we shall collect the basics of Bailey chains. The details and proofs can be found in \cite{Andrew:1984,Paule}.

A pair of sequences $(a_n,b_n)_{n\ge0}$ is called a Bailey pair if they are related by
\be
b_n=\sum_{j=0}^n\frac{a_j}{(q;q)_{n-j}(xq;q)_{n+j}}
\ee
or equivalently
\bea
a_n = \frac{1-xq^{2n}}{1-x}\sum_{j=0}^n(-1)^{n-j}q^{(n-j)(n-j-1)/2}\frac{(x;q)_{n+j}}{(q;q)_{n-j}}b_j \ .
\eea
Given a Bailey pair $(a_n,b_n)_{n\ge0}$, a new Baiely pair $(a'_n,b'_n)_{n\ge0}$ is constructed by 
\be\label{new pair}
a'_n=x^nq^{n^2} a_n \ \ \ \ b'_n=\sum_{j=0}^n\frac{x^jq^{j^2}}{(q;q)_{n-j}}b_j  \ \ \ n=0,1,2,3,\cdots 
\ee
A sequence $(a,b)\to (a',b')\to \cdots \to (a^{(\ell)},b^{(\ell)})\to \cdots $ obtained by applying \eqref{new pair} iteratively is called a \emph{Bailey chain}. It is known that two Bailey pairs next to each other in a Bailey chain, $(a_n^{(\ell-1)},b_n^{(\ell-1)})_{n\ge0}$ and $(a_n^{(\ell)},b_n^{(\ell)})_{n\ge0}$, obey the special version of the Bailey's lemma
\be\label{Bailey 2}
 \sum_{j=0}^n
    \frac{   a_j^{(\ell)}}{ (q;q)_{n-j} \, (x q;q)_{n+j}  }=\sum_{j=0}^n \frac{x^jq^{j^2} }{(q;q)_{n-j}}  \sum_{k=0}^j  \frac{  a_k^{(\ell-1)}}{ (q;q)_{j-k} \, (x q;q)_{j+k}} \ .
\ee
If we start with the simplest example of a Bailey pair \cite{Andrew:1984}
\bea
&&\a_n=(-1)^n q^{n(n-1)/2} \frac{1-x q^{2n}}{1-x}\frac{(x;q)_{n}}{(q;q)_{n}}  \ , \cr
&&\b_n=\delta_{n,0}  \ ,
\eea
then we can write the next pair in the Bailey chain as
\bea
&&\a'_n=(-1)^n x^n q^{n^2+n(n-1)/2} \frac{1-x q^{2n}}{1-x}\frac{(x;q)_{n}}{(q;q)_{n}}  \ , \cr
&&\b'_n=\frac1{(q;q)_n} \ .
\eea
By definition, the Bailey pair $(\a'_n,\b'_n)$ satisfies the identity 
\bea\label{Bailey 1}
\b'_n=\frac1{(q;q)_n}=\sum_{r=0}^{n}  \frac{  \a'_r }{ (q;q)_{{n}-{r}} \, (x q;q)_{n+r}} \ .
\eea
Using the special version of the Bailey's lemma \eqref{Bailey 2} recursively, we find
 \bea\label{recursive}
   && (q;q)_{s_{p}} \sum_{s_{p-1}=0}^{s_p}
    \frac{   \a_{s_{p-1}}^{(p)}\,   }{ (q;q)_{s_{p}-s_{p-1}} \, (x q;q)_{s_{p}+s_{p-1}}  }\cr 
    &\stackrel{\eqref{Bailey 2}}{=}&(q;q)_{s_{p}}\sum_{s_{p-1}=0}^{s_p} \frac{x^{s_{p-1}}q^{s_{p-1}^2} }{(q;q)_{n-s_{p-1}}}  \sum_{s_{p-2}=0}^{s_{p-1}}  \frac{ \a_{s_{p-2}}^{(p-1)} }{ (q;q)_{{s_{p-1}}-{s_{p-2}}} \, (x q;q)_{s_{p-1}+s_{p-2}}}\cr
    &\stackrel{\eqref{Bailey 2}}{=}&(q;q)_{s_{p}}\sum_{{s_p}\ge  \cdots\ge s_{1}\ge0} \prod_{i=1}^{p-1} x^{s_i} q^{s_i^2}\frac{1}{(q;q)_{s_p-s_{p-1}}\cdots (q;q)_{s_2-s_{1}} }\sum_{s_0=0}^{s_{1}}  \frac{  \a'_{s_0} }{ (q;q)_{{s_{1}}-{s_{0}}} \, (x q;q)_{s_{1}+s_{0}}}\cr
    &\stackrel{\eqref{Bailey 1}}{=}&\sum_{{s_p}\ge \cdots\ge s_{1}\ge0} \prod_{i=1}^{p-1} x^{s_i} q^{s_i^2}\frac{(q;q)_{s_{p}}}{(q;q)_{s_p-s_{p-1}}\cdots (q;q)_{s_2-s_{1}} (q;q)_{s_1}}\cr
    &=&\sum_{{s_p}\ge  \cdots\ge s_{1}\ge0} \prod_{i=1}^{p-1} x^{s_i} q^{s_i^2} \left[\begin{array}{c}s_{i+1}\\ s_{i} \end{array}\right]_q \ ,
  \eea
where 
\bea
\a_{n}^{(\ell)}=(-1)^n x^{\ell n} q^{ \ell n^2+n(n-1)/2} \frac{1-x q^{2n}}{1-x}\frac{(xq;q)_{n}}{(q;q)_{n}}   \ .
\eea
Once $x=at^2q^{-1}$ is substituted into \eqref{recursive}, the last line of \eqref{recursive} is exactly the same as the second line of \eqref{superpoly1}. Therefore,  by using the first line of \eqref{recursive} with a change of variables $s_{p}\to k$ and $s_{p-1}\to \ell$, we can manipulate the multi-sum expression \eqref{superpoly1} for $K_{p>0}$
\bea\label{superpoly5}
 \eqref{superpoly1}  &=& (-t)^{-n+1} \sum^\infty_{k=0}  \sum_{\ell=0}^k q^{k} \frac{(-a t q^{-1};q)_{k}}{(q;q)_{k}}  (q^{1-n};q)_{k} (-a t^3 q^{n-1};q)_{k} \cr
&&\times(-1)^\ell (a t^{2})^{p\ell} q^{p\ell(\ell-1)+\ell(\ell-1)/2} \frac{1-at^2 q^{2\ell-1}}{1-at^2q^{-1}}\frac{(at^2q^{-1};q)_{\ell}}{(q;q)_{\ell}} \frac{(q;q)_k}{(q;q)_{k-\ell}(at^2;q)_{k+\ell}}\cr
&=& (-t)^{-n+1} \sum^\infty_{k=0}  \sum_{\ell=0}^k q^{k} \frac{(-a t q^{-1};q)_{k}}{(q;q)_{k}}  (q^{1-n};q)_{k} (-a t^3 q^{n-1};q)_{k} \cr
&&\hspace{3cm}\times(-1)^\ell (a t^{2})^{p\ell}q^{(p+1/2)\ell(\ell-1)} \frac{1-at^2 q^{2\ell-1}}{(at^2q^{\ell-1};q)_{k+1}} \left[ \begin{array}{c} k \\ \ell \end{array}\right]_q \ .
\eea
In addition, it is straightforward to show the equivalence of \eqref{superpoly2} and \eqref{superpoly4} by using the fact \eqref{inverse}.

\section{Special colored superpolynomials}\label{sec:sp}
It was first observed in \cite{DuninBarkowski:2011yx} for several knots that the limit $q\to 1$ of the $n$-colored HOMFLY polynomial $P_n(K;a,q)$ which is called the special colored HOMFLY has the property
\be\label{special2}
\lim_{q\to 1 }P_n(K;a,q) = \Big[ \lim_{q\to 1 } P_{2}(K;a,q) \Big]^{n-1} \ . 
\ee
This property \eqref{special2} was proven for all knots in \cite{Zhu:2012tm}. Recently, it was proposed that this could be lifted to the level of colored superpolynomials \cite{Morozov:2012am}:
\be\label{special3}
\lim_{q\to 1 }\scP_n(K;a,q,t) = \Big[ \lim_{q\to 1 } \scP_{2}(K;a,q,t) \Big]^{n-1} \ . 
\ee
However, it is certainly \emph{not} satisfied by the knot $\bf 9_{42}$. (See appendix B in \cite{Gukov:2011ry}.) Therefore, this is not true in general.  It is believed that this property \eqref{special3} holds true for some simple knots such as $\scH$-thin knots, alternating knots and torus knots. In fact, the $n$-colored superpolynomials of the torus knots $T^{2,2p+1}$ and the figure-eight $\bf 4_1$ satisfy the property \eqref{special3}. 
This can be interpreted as the exponentially growth property at the homological level \cite{Gukov:2011ry}. In what follows, we will check \eqref{special3} in the case of  the ``special'' colored superpolynomials of the twist knots.

Using the expression \eqref{superpoly3}, the limit $q\to 1$ of the superpolynomials of $K_{p>0}$ can be written as
\be
\lim_{q\to 1 } \scP_{2}(K_{p>0};a,q,t)= (-t)^{-1}\left[1+(1+at) (1+at^3) (1-at^2)^{-1}(1-a^pt^{2p})\right] \ .
\ee
In this limit,  the colored superpolynomials of $K_{p>0}$ asymptotes to
\bea
&& \lim_{q\to 1 } \scP_{n}(K_{p>0};a,q,t)\cr&=&(-t)^{-n+1}\sum^{n-1}_{k=0} \left( \begin{array}{c} n-1\\ k\end{array}\right) (1+at)^{k} (1+at^3)^{k}(1-at^2)^{-k} \sum_{\ell=0}^k \left( \begin{array}{c}  k\\ \ell\end{array}\right)(-a^pt^{2p})^\ell \cr
 &=&(-t)^{-n+1}\sum_{k=0}^{n-1} \left( \begin{array}{c} n-1\\ k \end{array}\right)(1+at)^{k} (1+at^3)^{k} (1-at^2)^{-k}(1-a^pt^{2p} )^k\cr
 &=&(-t)^{-n+1}\left[1+(1+at) (1+at^3) (1-at^2)^{-1}(1-a^pt^{2p})\right] ^{n-1}\cr
 &=&\left[\lim_{q\to 1 } \scP_{2}(K_{p>0};a,q,t)\right]^{n-1} \ .
\eea
It is clear from the expression \eqref{superpoly4} that the ``special'' colored superpolynomial of  $K_{p<0}$ obeys the property  \eqref{special3}. 
Thus, we find that the proposal in \cite{Morozov:2012am} still holds in the case of the twist knots. However, it is still an open problem to classify knots which satisfy the exponential growth property \eqref{special3}.

\section{Colored HOMFLY polynomials of twist knots}\label{sec:10crossings}
\begin{table}[b]
\begin{center}
\begin{tabular}{|c|p{13cm}|}
\hline 
\textbf{ Knot} & $P_{2}(K;a,q)$\tabularnewline\hline 
\hline \rule{0pt}{5mm}
${\bf 4_1}$ & $a^{-1}-q^{-1}+1-q+a$\tabularnewline
\hline \rule{0pt}{5mm}
${\bf 5_2}$ & $a(q^{-1}-1+q)+a^{2}(q^{-1}-1+q)-a^{3}$\tabularnewline
\hline \rule{0pt}{5mm}
${\bf 6_1}$ & $a^{-2}+a^{-1}(-q^{-1}+1-q)-q^{-1}+2-q+a$\tabularnewline
\hline \rule{0pt}{5mm}
${\bf 7_2}$ & $a(q^{-1}-1+q)+a^{2}(q^{-1}-2+q)+a^{3}(q^{-1}-1+q)-a^{4}$\tabularnewline
\hline \rule{0pt}{5mm}
${\bf 8_1}$ & $a^{-3}+a^{-2}(-q^{-1}+1+q)+a^{-1}(-q^{-1}+2-q)-q^{-1}+2-q+a$\tabularnewline
\hline \rule{0pt}{5mm}
${\bf 9_2}$ & $a(q^{-1}-1+q)+a^{2}(q^{-1}-2+q)+a^{3}(q^{-1}-2+q)+a^{4}(q^{-1}-1+q)-a^{5}$\tabularnewline
\hline \rule{0pt}{5mm}
${\bf 10_1}$ & $a^{-4}+a^{-3}(-q^{-1}+1-q)+a^{-2}(-q^{-1}+2-q)+a^{-1}(-q^{-1}+2-q)-q^{-1}-q+2+a$\tabularnewline
\hline
\end{tabular}\caption{HOMFLY polynomials for the twist knots up to 10 crossings. Here we make a change of variables in the results obtained from \eqref{CS inv} such that $a\to a^{-1}$ and $q\to q^{-1}$.}
\label{tab:homfly}
\end{center}
\end{table}
\begin{table}
\begin{center}
\begin{tabular}{|c|p{13cm}|}
\hline 
\textbf{ Knot} & $P_{3}(K;a,q)$\tabularnewline
\hline 
\hline 
${\bf 4_1}$ & \footnotesize{$a^{-2}q^{-2}+a^{-1}(-q^{-3}-q^{-2}+q^{-1}-q)+(q^{-3}-q^{-2}+3-q^{2}+q^{3})$}\\  &
\footnotesize{$+a\left(-q^{-1}+q-q^{2}-q^{3}\right)+a^{2}q^{2}$}\tabularnewline
\hline
${\bf 5_2}$ & \footnotesize{$a^{2}(q^{-2}-q^{-1}-1+2q-q^{3}+q^{4})+a^{3}(q^{-2}+q^{-1}-2+3q^{2}-q^{3}-q^{4}+q^{5})$}\\  &
\footnotesize{$+a^{4}(-2q+2q^{3}-q^{4}-q^{5}+q^{6})+a^{5}(-q^{2}+q^{4}-q^{5}-q^{6})+a^{6}q^{5}$}\tabularnewline
\hline 
${\bf 6_1}$ & \footnotesize{$a^{-4}q^{-4}+a^{-3}(-q^{-5}-q^{-4}+q^{-3}-q^{-1})+a^{-2}(q^{-5}-q^{-4}-q^{-3}+2q^{-2}+q^{-1}-1)$}\\  &
\footnotesize{$+a^{-1}(q^{-4}-q^{-3}-2q^{-2}+3q^{-1}+2-2q+q^{3})+(q^{-3}-2q^{-2}-2q^{-1}+4-3q^{2})$}\\  &
\footnotesize{$+a\left(-q^{-1}+2q-q^{3}\right)+a^{2}q^{2}$}\tabularnewline
\hline 
${\bf 7_2}$ & \footnotesize{$a^{2}(q^{-2}-q^{-1}-1+2q-q^{3}+q^{4})+a^{3}(q^{-2}-3+q+3q^{2}-2q^{3}-q^{4}+q^{5})$}\\  &
\footnotesize{$+a^{4}(q^{-2}+q^{-1}-1-2q+2q^{2}+3q^{3}-2q^{4}-q^{5}+q^{6})$}\\  &
\footnotesize{$+a^{5}(-q-2q^{2}+q^{3}+2q^{4}-2q^{5}-q^{6}+q^{7})$}\\  &
\footnotesize{$+a^{6}(-q^{3}+q^{4}+2q^{5}-q^{6}-q^{7}+q^{8})+a^{7}(-q^{4}+q^{6}-q^{7}-q^{8})+a^{8}q^{7}$}\tabularnewline
\hline 
${\bf 8_1}$ & \footnotesize{$a^{-6}q^{-6}+a^{-5}(-q^{-7}-q^{-6}+q^{-5}-q^{-3})+a^{-4}(q^{-7}-q^{-6}-q^{-5}+2q^{-4}+q^{-3}-q^{-2})$}\\  &
\footnotesize{$+a^{-3}(q^{-6}-q^{-5}-2q^{-4}+2q^{-3}+q^{-2}-q^{-1})$}\\  &
\footnotesize{$+a^{-2}(q^{-5}-q^{-4}-2q^{-3}+2q^{-2}+2q^{-1}-q+q^{3})$}\\  &
\footnotesize{$+a^{-1}(3+q^{-4}-q^{-3}-3q^{-2}+2q^{-1}-2q-2q^{2})+4+q^{-3}-2q^{-2}-2q^{-1}+q-2q^{2}$}\\  &
\footnotesize{$a(-q^{-1}+2q-q^{3})+a^{2}q^{2}$}\tabularnewline
\hline
${\bf 9_2}$ & \footnotesize{$a^{2}(q^{-2}-q^{-1}-1+2q-q^{3}+q^{4})+a^{3}(q^{-2}-3+q+3q^{2}-2q^{3}-q^{4}+q^{5})$}
\\  &
\footnotesize{$+a^{4}(q^{-2}-2-q+2q^{2}+2q^{3}-2q^{4}-q^{5}+q^{6})$}\\  & 
\footnotesize{$+a^{5}(q^{-2}+q^{-1}-1-q+2q^{3}+2q^{4}-2q^{5}-q^{6}+q^{7})$}\\  &
\footnotesize{$+a^{6}(-q-q^{2}-q^{3}+q^{4}+2q^{5}-2q^{6}-q^{7}+q^{8})$}\\  &
\footnotesize{$+a^{7}(-q^{4}+q^{5}+2q^{6}-2q^{7}-q^{8}+q^{9})+a^{8}(-q^{5}+q^{6}+2q^{7}-q^{8}-q^{9}+q^{10})$}\\  &
\footnotesize{$+a^{9}(-q^{6}+q^{8}-q^{9}-q^{10})+a^{10}q^{9}$}\tabularnewline
\hline 
${\bf 10_1}$ & \footnotesize{$a^{-8}q^{-8}+a^{-7}(-q^{-9}-q^{-8}+q^{-7}-q^{-5})+a^{-6}(q^{-9}-q^{-8}-q^{-7}+2q^{-6}+q^{-5}-q^{-4})$}
\\  &
\footnotesize{$+a^{-5}(q^{-8}-q^{-7}-2q^{-6}+2q^{-5}+q^{-4}-q^{-3})+a^{-4}(q^{-7}-q^{-6}-2q^{-5}+2q^{-4}+q^{-3}-q^{-2})$}\\  &
\footnotesize{$+a^{-3}(1+q^{-6}-q^{-5}-2q^{-4}+2q^{-3}+q^{-2}-q+q^{3})$}\\  &
\footnotesize{$+a^{-2}(1+q^{-5}-q^{-4}-2q^{-3}+q^{-2}+q^{-1}-q-2q^{2})$}\\  &
\footnotesize{$a^{-1}(3+q^{-4}-q^{-3}-3q^{-2}+2q^{-1}-q-q^{2})+4+q^{-3}-2q^{-2}-2q^{-1}+q-2q^{2}
 $}
\\  &
\footnotesize{$a(-q^{-1}+2q-q^{3})+a^{2}q^{2}$}\tabularnewline
\hline 
\end{tabular}
\caption{Colored HOMFLY polynomials for the twist knots up to 10 crossings. Here we make a change of variables in the results obtained from \eqref{CS inv} such that $a\to a^{-1}$ and $q\to q^{-1}$. }
\label{tab:clrhomfly}
\end{center}
\end{table}
 In this appendix, we show the colored HOMFLY polynomials $P_n(K_p;a,q)$ of the twist knots up to $10$ crossings with $n=2,3$. 
 Within the framework of $SU(N)$  Chern-Simons theory, we could directly  write the invariant of  non-torus knots,
 the invariant involves  quantum Racah coefficients. Unlike  quantum $SU(2)$ Racah coefficients \cite{Kirillov:1989},  
we do not have a closed form expression for   quantum $SU(N)$ Racah coefficients. This is the 
 barrier preventing us from writing the explicit polynomial  form of $V_R(K;a=q^N,q)$. Nevertheless,  
 the colored HOMFLY polynomials of non-torus knots can be determined by the  quantum Racah coefficents \cite{RamaDevi:1992dh,Zodinmawia:2011ud}  available for few representations. (The same results were obtained by character expasions \cite{Itoyama:2012qt}.) We refer the reader to \cite{Zodinmawia:2011ud,Zodinmawia:2012sx} for more detail.

The knot invariant of the twist knot $K_{p}$ carrying an arbitrary $SU(N)$ representation $R$ is given by the field theoretic method mentioned in \S\ref{sec:intro}:
\begin{eqnarray}\label{CS inv}
 V_{R}^{SU(N)}(K_{p};q) & = & \sum_{s,s^{\prime}\in R\otimes\bar{R}}\epsilon_{s}^{R,\bar{R}}\,\sqrt{\dim_{q}s}\,\epsilon_{s^{\prime}}^{R,\bar{R}}\,\sqrt{\dim_{q}s^{\prime}}\,(\lambda_{s}^{(-)}(R,\,\bar{R}))^{-2}\cr
&&\hspace{3cm}\times a_{ss^{\prime}}\left[\footnotesize{\begin{array}{cc}
R & \bar{R}\\
R & \bar{R}
\end{array}}\right] (\lambda_{s^{\prime}}^{(-)}(R,\,\bar{R}))^{-2p} 
 \ .
\end{eqnarray}
In the above equation, we denote the quantum dimension of the
representation $s$ by $\dim_q s$, and the quantum $SU(N)$ Racah coefficient is denoted by $a_{ts}$. The braiding eigenvalues $\lambda$ for parallel and anti-parallel
strand in standard framing are given by
\begin{equation}
\lambda_{Q}^{(+)}(R,\, R)=\epsilon_{Q;R,R}^{(+)}q^{2C_{R}-C_{Q}/2},\quad\quad\lambda_{P}^{(-)}(R,\,\bar{R})=\epsilon_{P;R,\bar{R}}^{(-)}q^{C_{P}/2}
\end{equation}
where $C_R$ is the quadratic Casimir of  the representation $R$.
It turns out that the field theoretic invariant is equal to the unnormalized (unreduced) colored HOMFLY polynomial $\overline{P}_{R}(K_{p};a,q)$ by replacing $q^N$ by $a$ in $V_{R}^{SU(N)}(K_{p};q)$.
One can normalize the knot invariant $\overline{P}_{R}(K_{p};a,q)$ by
an unknot factor
\begin{equation}
P_{R}(K_p;a,q)=\frac{\overline{P}_{R}(K_p;a,q)}{\dim_{q}R} \ .
\end{equation}
Using the  quantum $SU(N)$ Racah coefficients proposed in \cite{Zodinmawia:2011ud}, the HOMFLY and colored HOMFLY polynomials with $n=3$ are evaluated for the twist knots up to 10 crossings and listed in Table \ref{tab:homfly} and Table \ref{tab:clrhomfly} (these results are also given in \cite{Zodinmawia:2012sx}). We have checked that the polynomials obtained by plugging $t=-1$ into \eqref{superpoly1} and \eqref{superpoly2} also give the same results.

\section{Ooguri-Vafa conjecture}\label{sec:OV} 

In this appendix, we will check the validity of the formulae \eqref{superpoly1} and \eqref{superpoly2} in the context of Ooguri-Vafa conjecture \cite{Ooguri:1999bv}.
Ooguri-Vafa polynomials (or reformulated invariants) $f_R(a, q)$ and unnormalized HOMFLY polynomials $\bar{P}(a, q)$ are related by the equality \citep{Ooguri:1999bv}
\be
{\rm{log}}\sum_{R}\bar{P}_{R}(a,q)\Tr_{R}V=\sum_{d=1}^{\infty}\sum_{R}\frac1df_{R}(a^d, q^d)\Tr_{R}V^{d}.
\ee
Using group theoretical method, it can be shown that \citep{Labastida:2001ts}
\begin{eqnarray}
f_{R}(a,q) & = & \sum_{d,m=1}^{\infty}(-1)^{m-1}\frac{\mu(d)}{dm}\sum_{\{\vec{k}_{j}, R_{j}\}}\chi_{R}\left(C\left((\sum_{j=1}^{m}\vec{k}_{j})_{d}\right)\right)\nonumber \\
 &  & \times\prod_{j=1}^{m}\frac{|C(\vec{k}_{j})|}{l_{j}!}\chi_{R_{j}}(C(\vec{k}_{j}))\bar{P}_{R_{j}}(a^{d}, q^{d})
\end{eqnarray}
where $\mu(d)$ is the Moebius function and the second sum runs over
all vectors $\vec{k}_j>0$ and over representations
$R_{j}$. Further $(\vec{k}_{j})_{d}$
is defined as follows: $(\vec{k}_d)_{di}=k_i$ and
has zero entries for the other components. Therefore, if $\vec{k}=(k_{1},k_{2},\ldots)$,
then
\begin{equation}
\vec{k}_{d}=(0,...,0,k_{1},0,...,0,k_{2},...)
\end{equation}
where $k_{1}$ is in the $d$-th entry, $k_{2}$ in the $2d-$th entry,
and so on. Here $C(\vec{k})$ denotes the conjugacy class
determined by the sequence $(k_{1},k_{2},\cdots)$ (i.e
there are $k_{1}$ 1-cycles, $k_{2}$ 2-cycles etc) in the permutation
group $\ensuremath{S_{\ell}}(\ell=\sum_{j}jk_{j})$.
For a Young Tableau representation $R$ with $\ell$ number of boxes,
$\chi_{R}(C(\vec{k}))$ gives the character of the conjugacy
class $C(\vec{k})$ in the representation $R$.

The Ooguri-Vafa conjecture \citep{Ooguri:1999bv} claims that the function $f_{R}(a, q)$
has the following structure 
\begin{equation}
f_{R}(a,q)=\sum_{s,Q}\frac{N_{Q,R,s}}{q^{1/2}-q^{-1/2}}a^{Q}q^{s}
\end{equation}
where $N_{Q,R,s}$ are integer and $Q$ and $s$ are, in general,
half integers. The first few OV polynomials in terms of colored HOMFLY polynomials are
\bea
f_{2}&=&\bar{P}_{2}\cr
f_{3}&=&\bar{P}_{3}-\frac{1}{2}\bar{P}_{2}^2-
\frac{1}{2}\bar{P}_{2}^{(2)}\cr
f_{4}&=&\bar{P}_{4}-\bar{P}_{2}\bar{P}_{3}+
\frac{1}{3}\bar{P}_{2}^3-
\frac{1}{3}\bar{P}_{2}^{(3)}\cr
f_{5}&=&\bar{P}_{5}-\bar{P}_{2}\bar{P}_{4}-
\frac{1}{2}\bar{P}_{3}^2+
\bar{P}_{2}^2\bar{P}_{3}-\frac{1}{4}\bar{P}_{2}^4
-\frac{1}{2}\bar{P}_{3}^{(2)}+\frac{1}{4}\left(\bar{P}_{2}^{(2)}\right)^2
\eea
where
$\bar{P}_R^{(n)}(a, q )\equiv \bar{P}_R(a^n, q^n)$. 
Using the unnormalized colored HOMFLY polynomials obtained using our conjecture we have calculated the OV polynomials of the twist knots up to 8 crossings with representations $n=4 \ (R=\yng(3))$ and $n=5\ (R=\yng(4))$  
and has verified that these polynomials indeed has the structure predicted by Ooguri-Vafa conjecture. 
For reference, we write the OV polynomials only for the knot ${\bf 5_2}$ and the results for the other twist knot $K_p$ $(|p| \leq 4)$ are presented in the ancillary file\footnote{We would like to thank Stavros Garoufalidis for pointing out our mistake in the previous version.}.

\begin{table}[b]
\begin{center}
\begin{tabular}{|c|p{12.5cm}|}
\hline 
$f_{4}({\bf 5_{2}};a,q)$ & \footnotesize{$\tfrac{a^{\frac{3}{2}}(a-1)^{2}(aq-1)(q-a)}{(q^{\frac{1}{2}}-q^{-\frac{1}{2}}) q}\Big[-q + 2 q^2 - q^4 + q^5 + a (-q + q^2 + q^3 - q^4) + 
 a^2 (-1 - q - q^2 - q^4 - q^5 - q^6) + 
 a^3 (-1 - q - q^2 - q^3 - q^4 - q^5 - q^6 - q^7) + 
 a^4 (-1 - q - q^2 - 2 q^3 - 2 q^4 - 2 q^5 - 2 q^6 - 3 q^7 - q^8 - 
    2 q^9 - q^{10} - q^{11} - q^{13}) + 
 a^5 (q^2 + q^3 + q^4 + 2 q^5 + 2 q^6 + q^7 + 2 q^8 + q^9 + q^{10} + 
    q^{11} + q^{12} + q^{14})
\Big]$}
 \tabularnewline
\hline 
$f_{5}({\bf 5_{2}};a,q)$ &\footnotesize{$ - \tfrac{a^{2}(a-1)^{2}(aq-1)(q-a)}{(q^{\frac{1}{2}}-q^{-\frac{1}{2}})q^{\frac{3}{2}}}\Bigl[q^{2}-2q^{3}+q^{4}+q^{5}+q^{7}-q^{8}+q^{9}
 +a(-q+q^{2}-q^{3}-q^{4}+q^{5}-q^{6}-q^{8}-q^{11})
+a^{3}(-2q^{2}-q^{4}-4q^{5}-2q^{6}
 -2q^{7}-2q^{8}-q^{9}-q^{10}-q^{11})+a^{2}(-q^{2}-2q^{4}-3q^{5}-q^{6}-2q^{7}-2q^{8}
-q^{9}
 -q^{10}-q^{11})+a^{4}(1+q+2q^{2}+3q^{3}+6q^{4}+3q^{5}+6q^{6}+4q^{7}+5q^{8}+3q^{9}
 +4q^{10}+q^{11}+2q^{12}+q^{14})
+a^{5}(1+q^{2}+q^{3}+2q^{4}+2q^{5}+q^{6}+q^{7}+q^{8}
 -q^{11}-2q^{11}-2q^{13}-2q^{14}-3q^{15}-3q^{16}-2q^{17}-2q^{18}
-2q^{19}-2q^{20}-q^{21}-q^{22}
 -q^{24})+a^{6}(1+2q^{2}+2q^{3}+5q^{4}+6q^{5}+10q^{6}+10q^{7}+15q^{8}+15q^{9}+18q^{10}
 +17q^{11}+19q^{12}+16q^{13}+18q^{14}+13q^{15}+14q^{16}+11q^{17}+11q^{18}+7q^{19}
 +8q^{20}+5q^{22}+5q^{22}+3q^{23}
+2q^{24}+q^{25}+2q^{26})+a^{7}(-q-2q^{2}-4q^{3}-7q^{4}
 -10q^{5}-14q^{6}-17q^{7}-21q^{8}-22q^{9}-24q^{10}-24q^{11}
-24q^{12}-22q^{13}-21q^{14}
 -20q^{15}-18q^{16}-15q^{17}-13q^{18}-12q^{19}-9q^{20}-8q^{21}-6q^{22}-5q^{23}-4q^{24}
-3q^{25}
 -q^{26}-2q^{27}-q^{28})+a^{8}(q^{3}+q^{4}+3q^{5}+3q^{6}+6q^{7}+5q^{8}+8q^{9}+6q^{10}+9q^{11}
 +6q^{12}+9q^{13}
+6q^{14}+8q^{15}+5q^{16}+7q^{17}+4q^{18}+5q^{19}+3q^{20}+4q^{21}+2q^{22}
 +3q^{23}+q^{24}+2q^{25}+q^{26}+q^{27}+q^{29}\Big]$}
 \tabularnewline
\hline 
 \end{tabular}
\caption{Ooguri-Vafa polynomial of  the knot ${\bf 5_2}$}
\label{tab:cl OV poly}
\end{center}
\end{table}

\section{Tables and Figures}\label{sec: toobig}

\begin{figure}[H]
\begin{center}
\includegraphics[scale=0.86]{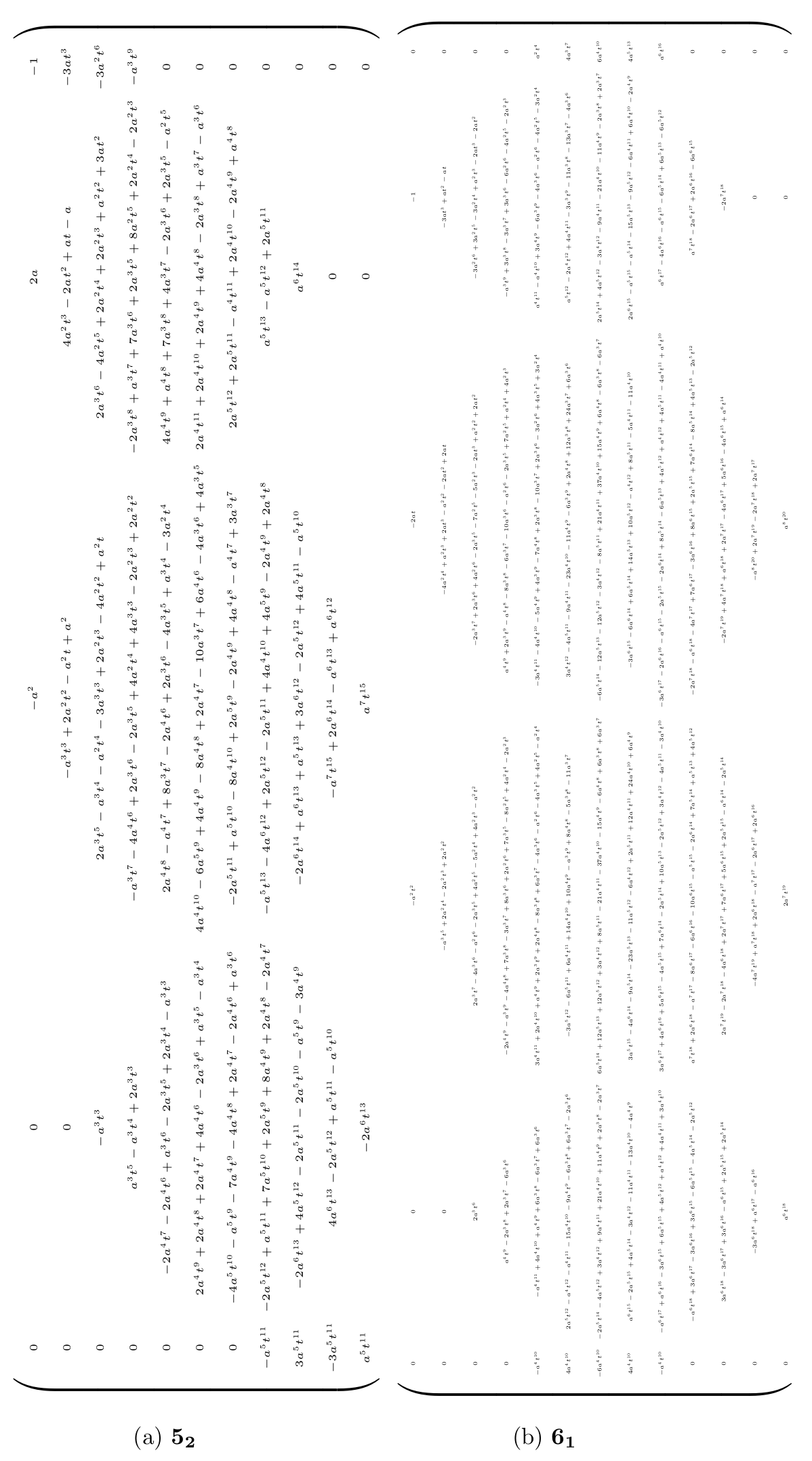} 
\caption{Matrix forms of classical super-$A$-polynomials for {$\bf 5_2$} and {$\bf 6_1$}. We normalized the results in Table \ref{tab:cl super a poly 1} and \ref{tab:cl super a poly 2} by the factors $(1+at^3x)^3$ and $a^2t^4x^4(1+at^3x)^4$ respectively.    Each entry in the $(i+1)^{\rm{th}}$ row and in the  $(j+1)^{\rm{th}}$ column gives the coefficient of $x^iy^j$.}\label{fig:mcap1}
\label{fig:matrixform}
\end{center}
\end{figure}

\begin{table}
\begin{center}
\begin{tabular}{|c|p{14cm}|}
\hline 
\textbf{Knot}& $A^{\rm super}(K;x,y;a,t)$ \tabularnewline
\hline 
\hline 
${\bf 6_1}$ & \footnotesize{$y^5$}\\[3pt] \rule{0pt}{2mm}
&$-\frac{1}{a^2 t^4 x^4 \left(1+a t^3 x\right)}$\footnotesize{ $(1 + a t x - a t^2 x + 2 a t^2 x^2 + 2 a t^3 x^2 - a^2 t^3 x^2 + 
 2 a^2 t^3 x^3 - 2 a^2 t^5 x^3 + 3 a^2 t^4 x^4 + 4 a^2 t^5 x^4 + 
 a^2 t^6 x^4 - 2 a^3 t^6 x^4 + 4 a^3 t^6 x^5 + 4 a^3 t^7 x^5 - 
 a^3 t^8 x^5 - 2 a^3 t^7 x^6 + 2 a^3 t^8 x^6 - a^4 t^9 x^6 + 
 2 a^4 t^9 x^7) y^4$}\\[3pt] \rule{0pt}{2mm}
&$+\frac{-1+x}{a t^3 x^4 \left(1+a t^3 x\right)^2}$\footnotesize{$(2 + 2 t x + a t x - 2 t^2 x - a t^2 x + 4 a t^2 x^2 + 4 a t^3 x^2 + 
 3 a t^4 x^2 + 3 a t^3 x^3 + 4 a^2 t^4 x^3 - 3 a t^5 x^3 + 
 6 a^2 t^5 x^4 + 12 a^2 t^6 x^4 + 6 a^2 t^7 x^4 - 6 a^2 t^6 x^5 - 
 6 a^2 t^7 x^5 + 6 a^3 t^7 x^5 + 3 a^3 t^8 x^5 + 2 a^3 t^9 x^6 + 
 3 a^3 t^{10} x^6 + a^3 t^9 x^7 - 4 a^3 t^{10} x^7 + 4 a^4 t^{10} x^7 + 
 a^3 t^{11} x^7 + 2 a^4 t^{11} x^7 - 2 a^4 t^{11} x^8 + 2 a^4 t^{12} x^8 + 
 a^5 t^{13} x^9) y^3$}\\[3pt] \rule{0pt}{2mm}
&$+\frac{(-1+x)^2}{t^2 x^4 \left(1+a t^3 x\right)^3}$\footnotesize{$ (-1 - 2 t x + 2 t^2 x - t^2 x^2 + 4 t^3 x^2 - 4 a t^3 x^2 - t^4 x^2 - 
 2 a t^4 x^2 + 2 a t^5 x^3 + 3 a t^6 x^3 + 6 a t^5 x^4 + 
 6 a t^6 x^4 - 6 a^2 t^6 x^4 - 3 a^2 t^7 x^4 + 6 a^2 t^7 x^5 + 
 12 a^2 t^8 x^5 + 6 a^2 t^9 x^5 - 3 a^2 t^8 x^6 - 4 a^3 t^9 x^6 + 
 3 a^2 t^{10} x^6 + 4 a^3 t^{10} x^7 + 4 a^3 t^{11} x^7 + 3 a^3 t^12 x^7 - 
 2 a^3 t^{12} x^8 - a^4 t^{12} x^8 + 2 a^3 t^{13} x^8 + a^4 t^{13} x^8 + 
 2 a^4 t^{14} x^9) y^2$}\\[3pt] \rule{0pt}{2mm}
&$+\frac{a t^2 (-1+x)^3}{x^2 \left(1+a t^3 x\right)^4}$\footnotesize{$(-2 - 2 t x + 2 t^2 x - a t^3 x - 4 a t^3 x^2 - 4 a t^4 x^2 + 
 a t^5 x^2 + 3 a t^4 x^3 + 4 a t^5 x^3 + a t^6 x^3 - 2 a^2 t^6 x^3 - 
 2 a^2 t^6 x^4 + 2 a^2 t^8 x^4 + 2 a^2 t^8 x^5 + 2 a^2 t^9 x^5 - 
 a^3 t^9 x^5 - a^3 t^{10} x^6 + a^3 t^{11} x^6 + a^3 t^12 x^7)y$}\\[3pt] \rule{0pt}{2mm}
&$-\frac{a^2 t^6 (-1+x)^4}{\left(1+a t^3 x\right)^4}$
\tabularnewline
\hline
${\bf 7_2}$ &  \footnotesize{$y^6$}\\[3pt] \rule{0pt}{2mm}
&$-\frac{a}{1+a t^3 x}$\footnotesize{$(3-2 x+2 t x-3 t^2 x-t x^2+4 t^2 x^2+2 a t^2 x^2+6 a t^3 x^2-a t^2 x^3-3 a t^3 x^3+4 a t^4 x^3+3 a t^5 x^3-2 a t^4 x^4+a^2 t^4 x^4-2 a t^5 x^4+4 a^2 t^5 x^4+3 a^2 t^6 x^4-2 a^2 t^5 x^5+2 a^2 t^7 x^5-a^2 t^7 x^6+2 a^3 t^7 x^6+2 a^3 t^8 x^6-a^3 t^8 x^7+a^3 t^9 x^7+a^4 t^{10} x^8) y^5$}\\[3pt]\rule{0pt}{5mm}
&$-\frac{a^2 (-1+x)}{\left(1+a t^3 x\right)^2}$\footnotesize{$(3-x+4 t x-6 t^2 x-t x^2+7 t^2 x^2+2 a t^2 x^2-6 t^3 x^2+12 a t^3 x^2+3 t^4 x^2+4 t^3 x^3-2 a t^3 x^3-6 t^4 x^3+8 a t^4 x^3-6 a t^5 x^3+a t^4 x^4+9 a t^5 x^4+8 a^2 t^5 x^4-10 a t^6 x^4+18 a^2 t^6 x^4-9 a t^7 x^4+8 a t^6 x^5-4 a^2 t^6 x^5+8 a t^7 x^5+8 a^2 t^7 x^5+6 a^2 t^8 x^5+5 a^2 t^7 x^6-3 a^2 t^8 x^6+12 a^3 t^8 x^6-2 a^2 t^9 x^6+12 a^3 t^9 x^6+3 a^2 t^{10} x^6+4 a^2 t^9 x^7-6 a^3 t^9 x^7-a^2 t^{10} x^7+8 a^3 t^{10} x^7+6 a^3 t^{11} x^7+3 a^3 t^{10} x^8-5 a^3 t^{11} x^8+8 a^4 t^{11} x^8+2 a^3 t^{12} x^8+3 a^4 t^{12} x^8-3 a^4 t^{12} x^9+4 a^4 t^{13} x^9+2 a^5 t^{14} x^{10}) y^4$}\\[3pt]\rule{0pt}{2mm}
&$-\frac{a^3 (-1+x)^2}{\left(1+a t^3 x\right)^3}$\footnotesize{$(1+2 t x-3 t^2 x+3 t^2 x^2-8 t^3 x^2+6 a t^3 x^2+3 t^4 x^2+3 t^3 x^3-9 t^4 x^3+6 a t^4 x^3+6 t^5 x^3-9 a t^5 x^3-t^6 x^3-6 t^5 x^4+6 a t^5 x^4+4 t^6 x^4-20 a t^6 x^4+15 a^2 t^6 x^4-6 a t^7 x^4+3 a t^6 x^5-9 a t^7 x^5+4 a^2 t^7 x^5+6 a t^8 x^5-6 a^2 t^8 x^5+9 a t^9 x^5-12 a t^8 x^6+6 a^2 t^8 x^6-12 a t^9 x^6-24 a^2 t^9 x^6+20 a^3 t^9 x^6-18 a^2 t^{10} x^6-3 a^2 t^9 x^7+9 a^2 t^{10} x^7-4 a^3 t^{10} x^7-6 a^2 t^{11} x^7+6 a^3 t^{11} x^7-9 a^2 t^{12} x^7-6 a^2 t^{11} x^8+6 a^3 t^{11} x^8+4 a^2 t^{12} x^8-20 a^3 t^{12} x^8+15 a^4 t^{12} x^8-6 a^3 t^{13} x^8-3 a^3 t^{12} x^9+9 a^3 t^{13} x^9-6 a^4 t^{13} x^9-6 a^3 t^{14} x^9+9 a^4 t^{14} x^9+a^3 t^{15} x^9+3 a^4 t^{14} x^{10}-8 a^4 t^{15} x^{10}+6 a^5 t^{15} x^{10}+3 a^4 t^{16} x^{10}-2 a^5 t^{16} x^{11}+3 a^5 t^{17} x^{11}+a^6 t^{18} x^{12}) y^3$}\\[3pt] \rule{0pt}{2mm}
&$+\frac{a^4 t^3 (-1+x)^3 x^2}{\left(1+a t^3 x\right)^5}$\footnotesize{$(2+3 t x-4 t^2 x+3 t^2 x^2-5 t^3 x^2+8 a t^3 x^2+2 t^4 x^2+3 a t^4 x^2-4 t^4 x^3+6 a t^4 x^3+t^5 x^3-8 a t^5 x^3-6 a t^6 x^3+5 a t^5 x^4-3 a t^6 x^4+12 a^2 t^6 x^4-2 a t^7 x^4+12 a^2 t^7 x^4+3 a t^8 x^4-8 a t^7 x^5+4 a^2 t^7 x^5-8 a t^8 x^5-8 a^2 t^8 x^5-6 a^2 t^9 x^5+a^2 t^8 x^6+9 a^2 t^9 x^6+8 a^3 t^9 x^6-10 a^2 t^{10} x^6+18 a^3 t^{10} x^6-9 a^2 t^{11} x^6-4 a^2 t^{10} x^7+2 a^3 t^{10} x^7+6 a^2 t^{11} x^7-8 a^3 t^{11} x^7+6 a^3 t^{12} x^7-a^3 t^{11} x^8+7 a^3 t^{12} x^8+2 a^4 t^{12} x^8-6 a^3 t^{13} x^8+12 a^4 t^{13} x^8+3 a^3 t^{14} x^8+a^4 t^{13} x^9-4 a^4 t^{14} x^9+6 a^4 t^{15} x^9+3 a^5 t^{16} x^{10})y^2$}\\[3pt] \rule{0pt}{2mm}
&$-\frac{a^5 t^6 (-1+x)^4 x^4}{\left(1+a t^3 x\right)^5}$\footnotesize{$(1+t x-t^2 x-t^3 x^2+2 a t^3 x^2+2 a t^4 x^2+2 a t^4 x^3-2 a t^6 x^3-2 a t^6 x^4+a^2 t^6 x^4-2 a t^7 x^4+4 a^2 t^7 x^4+3 a^2 t^8 x^4+a^2 t^7 x^5+3 a^2 t^8 x^5-4 a^2 t^9 x^5-3 a^2 t^{10} x^5-a^2 t^9 x^6+4 a^2 t^{10} x^6+2 a^3 t^{10} x^6+6 a^3 t^{11} x^6+2 a^3 t^{11} x^7-2 a^3 t^{12} x^7+3 a^3 t^{13} x^7+3 a^4 t^{14} x^8) y$}\\[3pt] \rule{0pt}{2mm}
&$+\frac{a^8 t^{18} (-1+x)^5 x^{11}}{\left(1+a t^3 x\right)^5}$
\tabularnewline
\hline 
 \end{tabular}
\caption{Classical super-$A$-polynomial of the knots ${\bf 6_1}$ and ${\bf 7_2}$}
\label{tab:cl super a poly 2}
\end{center}
\end{table}

\begin{table}[H]
\begin{center}
\begin{tabular}{|c|p{12.8cm}|}
\hline 
\textbf{Knot}& $A^{\rm super}(K;x,y;a,t)$ \tabularnewline
\hline 
\hline 
${\bf 8_1}$ &\footnotesize{$y^7$}\\[3pt] \rule{0pt}{2mm}
&$-\frac{1}{a^3 t^6 x^6 (1+a t^3 x)}$\footnotesize{$(1+a t x-a t^2 x+2 a t^2 x^2+2 a t^3 x^2-a^2 t^3 x^2+2 a^2 t^3 x^3-2 a^2 t^5 x^3+3 a^2 t^4 x^4+4 a^2 t^5 x^4-2 a^3 t^5 x^4+a^2 t^6 x^4-2 a^3 t^6 x^4+3 a^3 t^5 x^5+a^3 t^6 x^5-3 a^3 t^7 x^5-a^3 t^8 x^5+4 a^3 t^6 x^6+6 a^3 t^7 x^6+2 a^3 t^8 x^6-4 a^4 t^8 x^6-a^4 t^9 x^6+6 a^4 t^8 x^7+6 a^4 t^9 x^7-2 a^4 t^{10} x^7-3 a^4 t^9 x^8+3 a^4 t^{10} x^8-2 a^5 t^{11} x^8+3 a^5 t^{11} x^9) y^6$}\\[3pt] \rule{0pt}{2mm}
&$+\frac{-1+x}{a^2 t^5 x^6 (1+a t^3 x)^2}$\footnotesize{$(3+3 t x+2 a t x-3 t^2 x-2 a t^2 x+8 a t^2 x^2+6 a t^3 x^2-a^2 t^3 x^2+4 a t^4 x^2+6 a t^3 x^3+3 a^2 t^3 x^3+2 a t^4 x^3+4 a^2 t^4 x^3-4 a t^5 x^3-3 a^2 t^5 x^3+9 a^2 t^4 x^4+17 a^2 t^5 x^4+15 a^2 t^6 x^4-4 a^3 t^6 x^4+7 a^2 t^7 x^4+6 a^2 t^5 x^5+3 a^2 t^6 x^5+12 a^3 t^6 x^5-2 a^2 t^7 x^5+10 a^3 t^7 x^5+a^2 t^8 x^5-2 a^3 t^8 x^5+12 a^3 t^7 x^6+34 a^3 t^8 x^6+24 a^3 t^9 x^6-6 a^4 t^9 x^6+2 a^3 t^{10} x^6-12 a^3 t^8 x^7-12 a^3 t^9 x^7+18 a^4 t^9 x^7+2 a^3 t^{10} x^7+16 a^4 t^{10} x^7+2 a^3 t^{11} x^7-3 a^4 t^{11} x^7-3 a^4 t^{10} x^8+10 a^4 t^{11} x^8+12 a^4 t^{12} x^8-4 a^5 t^{12} x^8-a^4 t^{13} x^8+3 a^4 t^{11} x^9-9 a^4 t^{12} x^9+12 a^5 t^{12} x^9+3 a^4 t^{13} x^9+8 a^5 t^{13} x^9-2 a^5 t^{14} x^9-6 a^5 t^{13} x^{10}+6 a^5 t^{14} x^{10}-a^6 t^{15} x^{10}+3 a^6 t^{15} x^{11}) y^5$}\\[3pt] \rule{0pt}{2mm}
&$-\frac{(-1+x)^2}{a t^4 x^6 (1+a t^3 x)^3}$\footnotesize{$(3+6 t x+a t x-6 t^2 x-a t^2 x+3 t^2 x^2+6 a t^2 x^2-9 t^3 x^2+10 a t^3 x^2+3 t^4 x^2+6 a t^4 x^2+9 a t^3 x^3+3 a t^4 x^3+6 a^2 t^4 x^3-7 a t^5 x^3-2 a^2 t^5 x^3-6 a t^6 x^3+4 a t^4 x^4-12 a t^5 x^4+21 a^2 t^5 x^4-16 a t^6 x^4+29 a^2 t^6 x^4+14 a^2 t^7 x^4-4 a^2 t^7 x^5+15 a^3 t^7 x^5-12 a^2 t^8 x^5+2 a^3 t^8 x^5-8 a^2 t^9 x^5-18 a^2 t^7 x^6-24 a^2 t^8 x^6+24 a^3 t^8 x^6-14 a^2 t^9 x^6+40 a^3 t^9 x^6-8 a^2 t^{10} x^6+16 a^3 t^{10} x^6-24 a^3 t^9 x^7-50 a^3 t^{10} x^7+20 a^4 t^{10} x^7-24 a^3 t^{11} x^7+8 a^4 t^{11} x^7+2 a^3 t^{12} x^7+12 a^3 t^{10} x^8+6 a^4 t^{11} x^8-16 a^3 t^{12} x^8+21 a^4 t^{12} x^8-4 a^3 t^{13} x^8+14 a^4 t^{13} x^8-12 a^4 t^{12} x^9-26 a^4 t^{13} x^9+15 a^5 t^{13} x^9-10 a^4 t^{14} x^9+7 a^5 t^{14} x^9+4 a^4 t^{15} x^9-a^4 t^{13} x^{10}+9 a^4 t^{14} x^{10}-6 a^5 t^{14} x^{10}-9 a^4 t^{15} x^{10}+6 a^5 t^{15} x^{10}+a^4 t^{16} x^{10}+6 a^5 t^{16} x^{10}+3 a^5 t^{15} x^{11}-12 a^5 t^{16} x^{11}+6 a^6 t^{16} x^{11}+3 a^5 t^{17} x^{11}+2 a^6 t^{17} x^{11}-3 a^6 t^{17} x^{12}+3 a^6 t^{18} x^{12}+a^7 t^{19} x^{13}) y^4$}\\[3pt] \rule{0pt}{2mm}
&$-\frac{(-1+x)^3}{t^3 x^6 (1+a t^3 x)^4}$\footnotesize{$(-1-3 t x+3 t^2 x-3 t^2 x^2+12 t^3 x^2-6 a t^3 x^2-3 t^4 x^2-2 a t^4 x^2-t^3 x^3+9 t^4 x^3-6 a t^4 x^3-9 t^5 x^3+6 a t^5 x^3+t^6 x^3+6 a t^6 x^3+12 a t^5 x^4+26 a t^6 x^4-15 a^2 t^6 x^4+10 a t^7 x^4-7 a^2 t^7 x^4-4 a t^8 x^4+12 a t^6 x^5+6 a^2 t^7 x^5-16 a t^8 x^5+21 a^2 t^8 x^5-4 a t^9 x^5+14 a^2 t^9 x^5+24 a^2 t^8 x^6+50 a^2 t^9 x^6-20 a^3 t^9 x^6+24 a^2 t^{10} x^6-8 a^3 t^{10} x^6-2 a^2 t^{11} x^6-18 a^2 t^9 x^7-24 a^2 t^{10} x^7+24 a^3 t^{10} x^7-14 a^2 t^{11} x^7+40 a^3 t^{11} x^7-8 a^2 t^{12} x^7+16 a^3 t^{12} x^7+4 a^3 t^{12} x^8-15 a^4 t^{12} x^8+12 a^3 t^{13} x^8-2 a^4 t^{13} x^8+8 a^3 t^{14} x^8+4 a^3 t^{12} x^9-12 a^3 t^{13} x^9+21 a^4 t^{13} x^9-16 a^3 t^{14} x^9+29 a^4 t^{14} x^9+14 a^4 t^{15} x^9-9 a^4 t^{14} x^{10}-3 a^4 t^{15} x^{10}-6 a^5 t^{15} x^{10}+7 a^4 t^{16} x^{10}+2 a^5 t^{16} x^{10}+6 a^4 t^{17} x^{10}+3 a^4 t^{16} x^{11}+6 a^5 t^{16} x^{11}-9 a^4 t^{17} x^{11}+10 a^5 t^{17} x^{11}+3 a^4 t^{18} x^{11}+6 a^5 t^{18} x^{11}-6 a^5 t^{18} x^{12}-a^6 t^{18} x^{12}+6 a^5 t^{19} x^{12}+a^6 t^{19} x^{12}+3 a^6 t^{20} x^{13}) y^3$}\\[3pt] \rule{0pt}{2mm}
&$-\frac{a t (-1+x)^4}{x^4(1+a t^3 x)^5}$\footnotesize{$(-3-6 t x+6 t^2 x-a t^3 x-3 t^2 x^2+9 t^3 x^2-12 a t^3 x^2-3 t^4 x^2-8 a t^4 x^2+2 a t^5 x^2-3 a t^4 x^3+10 a t^5 x^3+12 a t^6 x^3-4 a^2 t^6 x^3-a t^7 x^3+12 a t^5 x^4+12 a t^6 x^4-18 a^2 t^6 x^4-2 a t^7 x^4-16 a^2 t^7 x^4-2 a t^8 x^4+3 a^2 t^8 x^4+12 a^2 t^7 x^5+34 a^2 t^8 x^5+24 a^2 t^9 x^5-6 a^3 t^9 x^5+2 a^2 t^{10} x^5-6 a^2 t^8 x^6-3 a^2 t^9 x^6-12 a^3 t^9 x^6+2 a^2 t^{10} x^6-10 a^3 t^{10} x^6-a^2 t^{11} x^6+2 a^3 t^{11} x^6+9 a^3 t^{10} x^7+17 a^3 t^{11} x^7+15 a^3 t^{12} x^7-4 a^4 t^{12} x^7+7 a^3 t^{13} x^7-6 a^3 t^{12} x^8-3 a^4 t^{12} x^8-2 a^3 t^{13} x^8-4 a^4 t^{13} x^8+4 a^3 t^{14} x^8+3 a^4 t^{14} x^8+8 a^4 t^{14} x^9+6 a^4 t^{15} x^9-a^5 t^{15} x^9+4 a^4 t^{16} x^9-3 a^4 t^{16} x^{10}-2 a^5 t^{16} x^{10}+3 a^4 t^{17} x^{10}+2 a^5 t^{17} x^{10}+3 a^5 t^{18} x^{11}) y^2$}\\[3pt] \rule{0pt}{2mm}
&$-\frac{a^2 t^5 (-1+x)^5}{x^2 (1+a t^3 x)^6}$\footnotesize{$(-3-3 t x+3 t^2 x-2 a t^3 x-6 a t^3 x^2-6 a t^4 x^2+2 a t^5 x^2+4 a t^4 x^3+6 a t^5 x^3+2 a t^6 x^3-4 a^2 t^6 x^3-a^2 t^7 x^3-3 a^2 t^6 x^4-a^2 t^7 x^4+3 a^2 t^8 x^4+a^2 t^9 x^4+3 a^2 t^8 x^5+4 a^2 t^9 x^5-2 a^3 t^9 x^5+a^2 t^{10} x^5-2 a^3 t^{10} x^5-2 a^3 t^{10} x^6+2 a^3 t^{12} x^6+2 a^3 t^{12} x^7+2 a^3 t^{13} x^7-a^4 t^{13} x^7-a^4 t^{14} x^8+a^4 t^{15} x^8+a^4 t^{16} x^9) y$}\\[3pt] \rule{0pt}{2mm}
&$+\frac{a^3 t^9 (-1+x)^6}{(1+a t^3 x)^6}$
\tabularnewline
\hline 
 \end{tabular}
\caption{Classical super-$A$-polynomial of the knot ${\bf 8_1}$}
\label{tab:cl super a poly 3}
\end{center}
\end{table}

\begin{table}[H]
\begin{center}
\begin{tabular}{|c|p{14cm}|}
\hline 
\textbf{Knot}& $\hat{A}^{\rm super}(K;\hat{x},\hat{y};a,q,t)$ \tabularnewline
\hline 

${\bf 6_1}$ &  \footnotesize {$a^2 q^{-58}t^4\hat{x}^4 (q^2+a t^3 \hat{x})(q^3+a t^3 \hat{x})(q^4+a t^3 \hat{x})(q^5+a t^3 \hat{x})(q^8+a t^3 \hat{x}^2)(q^9+a t^3 \hat{x}^2)(q^{10}+a t^3 \hat{x}^2)(q^{11}+a t^3 \hat{x}^2) \hat{y}^5$}\\[3pt]\rule{0pt}{2mm}
&\footnotesize{$-q^{-45}(q^2+a t^3 \hat{x})(q^3+a t^3 \hat{x})(q^4+a t^3 \hat{x})(q^2+a t^3 \hat{x}^2)(q^7+a t^3 \hat{x}^2)(q^8+a t^3 \hat{x}^2)(q^9+a t^3 \hat{x}^2)(q^{12}+a q^{10} t \hat{x}-a q^{10} t^2 \hat{x}+a q^9 t^2 \hat{x}^2+a q^{10} t^2 \hat{x}^2+a q^6 t^3 \hat{x}^2-a^2 q^8 t^3 \hat{x}^2+a q^{11} t^3 \hat{x}^2+a^2 q^7 t^3 \hat{x}^3+a^2 q^8 t^3 \hat{x}^3+a^2 q^4 t^4 \hat{x}^3-a^2 q^7 t^4 \hat{x}^3-a^2 q^8 t^4 \hat{x}^3+a^2 q^9 t^4 \hat{x}^3-a^2 q^4 t^5 \hat{x}^3-a^2 q^9 t^5 \hat{x}^3+a^2 q^6 t^4 \hat{x}^4+a^2 q^7 t^4 \hat{x}^4+a^2 q^8 t^4 \hat{x}^4+a^2 q^3 t^5 \hat{x}^4+a^2 q^4 t^5 \hat{x}^4+a^2 q^8 t^5 \hat{x}^4+a^2 q^9 t^5 \hat{x}^4-a^3 q^2 t^6 \hat{x}^4+a^2 q^5 t^6 \hat{x}^4-a^3 q^7 t^6 \hat{x}^4+a^3 q t^6 \hat{x}^5+a^3 q^2 t^6 \hat{x}^5+a^3 q^6 t^6 \hat{x}^5+a^3 q^7 t^6 \hat{x}^5+2 a^3 q^3 t^7 \hat{x}^5+a^3 q^4 t^7 \hat{x}^5+a^3 q^5 t^7 \hat{x}^5-a^3 q^3 t^8 \hat{x}^5-a^3 q^3 t^7 \hat{x}^6-a^3 q^4 t^7 \hat{x}^6+a^3 q^2 t^8 \hat{x}^6+a^3 q^3 t^8 \hat{x}^6-a^4 q t^9 \hat{x}^6+a^4 t^9 \hat{x}^7+a^4 q t^9 \hat{x}^7) \hat{y}^4$}\\[3pt] \rule{0pt}{2mm}
&\footnotesize{$+a q^{-30} t (-1+\hat{x})(q^2+a t^3 \hat{x})(q^3+a t^3 \hat{x})(q^2+a t^3 \hat{x}^2)(q^6+a t^3 \hat{x}^2)(q^7+a t^3 \hat{x}^2) (1+a q t^3 \hat{x}^2)(q^{11}+q^{12}+a q^{10} t \hat{x}+q^{11} t \hat{x}+q^{12} t \hat{x}-q^{10} t^2 \hat{x}-a q^{10} t^2 \hat{x}-q^{11} t^2 \hat{x}+a q^9 t^2 \hat{x}^2+2 a q^{10} t^2 \hat{x}^2+a q^{11} t^2 \hat{x}^2+a q^6 t^3 \hat{x}^2+2 a q^7 t^3 \hat{x}^2-a q^9 t^3 \hat{x}^2-a q^{10} t^3 \hat{x}^2+2 a q^{11} t^3 \hat{x}^2+a q^{12} t^3 \hat{x}^2+a q^8 t^4 \hat{x}^2+a q^9 t^4 \hat{x}^2+a q^{10} t^4 \hat{x}^2+a q^9 t^3 \hat{x}^3+a q^{10} t^3 \hat{x}^3+a q^{11} t^3 \hat{x}^3+a^2 q^5 t^4 \hat{x}^3+a q^6 t^4 \hat{x}^3+a^2 q^6 t^4 \hat{x}^3+a q^7 t^4 \hat{x}^3-a q^8 t^4 \hat{x}^3-2 a q^9 t^4 \hat{x}^3+a^2 q^9 t^4 \hat{x}^3-a q^{10} t^4 \hat{x}^3+a^2 q^{10} t^4 \hat{x}^3+a q^{11} t^4 \hat{x}^3+a q^{12} t^4 \hat{x}^3-a q^5 t^5 \hat{x}^3-a^2 q^5 t^5 \hat{x}^3-a q^6 t^5 \hat{x}^3+a^2 q^7 t^5 \hat{x}^3+a q^8 t^5 \hat{x}^3+a^2 q^8 t^5 \hat{x}^3-a q^{10} t^5 \hat{x}^3-a^2 q^{10} t^5 \hat{x}^3-a q^{11} t^5 \hat{x}^3+a^2 q^4 t^5 \hat{x}^4+2 a^2 q^5 t^5 \hat{x}^4+a^2 q^6 t^5 \hat{x}^4-a^2 q^7 t^5 \hat{x}^4-a^2 q^8 t^5 \hat{x}^4+a^2 q^9 t^5 \hat{x}^4+2 a^2 q^{10} t^5 \hat{x}^4+a^2 q^{11} t^5 \hat{x}^4+a^2 q^2 t^6 \hat{x}^4-a^2 q^4 t^6 \hat{x}^4+4 a^2 q^6 t^6 \hat{x}^4+5 a^2 q^7 t^6 \hat{x}^4+2 a^2 q^8 t^6 \hat{x}^4+a^2 q^{11} t^6 \hat{x}^4+a^2 q^3 t^7 \hat{x}^4+a^2 q^4 t^7 \hat{x}^4+a^2 q^5 t^7 \hat{x}^4+a^2 q^8 t^7 \hat{x}^4+a^2 q^9 t^7 \hat{x}^4+a^2 q^{10} t^7 \hat{x}^4-a^2 q^6 t^6 \hat{x}^5-2 a^2 q^7 t^6 \hat{x}^5-2 a^2 q^8 t^6 \hat{x}^5-a^2 q^9 t^6 \hat{x}^5+a^3 q t^7 \hat{x}^5-a^2 q^3 t^7 \hat{x}^5-2 a^2 q^4 t^7 \hat{x}^5+a^3 q^4 t^7 \hat{x}^5-a^2 q^5 t^7 \hat{x}^5+2 a^3 q^5 t^7 \hat{x}^5+a^2 q^6 t^7 \hat{x}^5+a^3 q^6 t^7 \hat{x}^5+a^2 q^7 t^7 \hat{x}^5-a^2 q^8 t^7 \hat{x}^5-2 a^2 q^9 t^7 \hat{x}^5+a^3 q^9 t^7 \hat{x}^5-a^2 q^{10} t^7 \hat{x}^5+a^3 q^2 t^8 \hat{x}^5+a^2 q^3 t^8 \hat{x}^5+a^3 q^3 t^8 \hat{x}^5-a^2 q^5 t^8 \hat{x}^5-a^3 q^5 t^8 \hat{x}^5-a^2 q^6 t^8 \hat{x}^5+a^3 q^7 t^8 \hat{x}^5+a^2 q^8 t^8 \hat{x}^5+a^3 q^8 t^8 \hat{x}^5-a^3 q^2 t^8 \hat{x}^6-a^3 q^3 t^8 \hat{x}^6+a^3 q^4 t^8 \hat{x}^6+2 a^3 q^5 t^8 \hat{x}^6+a^3 q^6 t^8 \hat{x}^6-a^3 q^7 t^8 \hat{x}^6-a^3 q^8 t^8 \hat{x}^6+a^3 q t^9 \hat{x}^6+2 a^3 q^2 t^9 \hat{x}^6-2 a^3 q^4 t^9 \hat{x}^6-a^3 q^5 t^9 \hat{x}^6+a^3 q^6 t^9 \hat{x}^6+a^3 q^7 t^9 \hat{x}^6+a^3 q^3 t^{10} \hat{x}^6+a^3 q^4 t^{10} \hat{x}^6+a^3 q^5 t^{10} \hat{x}^6+a^3 q^5 t^9 \hat{x}^7+a^4 t^{10} \hat{x}^7+a^4 q t^{10} \hat{x}^7-a^3 q^3 t^{10} \hat{x}^7-2 a^3 q^4 t^{10} \hat{x}^7+a^4 q^4 t^{10} \hat{x}^7-a^3 q^5 t^{10} \hat{x}^7+a^4 q^5 t^{10} \hat{x}^7+a^4 q^2 t^{11} \hat{x}^7+a^3 q^3 t^{11} \hat{x}^7+a^4 q^3 t^{11} \hat{x}^7-a^4 q^2 t^{11} \hat{x}^8-a^4 q^3 t^{11} \hat{x}^8+a^4 q t^{12} \hat{x}^8+a^4 q^2 t^{12} \hat{x}^8+a^5 t^{13} \hat{x}^9) \hat{y}^3$}\\[3pt] \rule{0pt}{2mm}
&\footnotesize$+a^2 q^{-16} t^2 (-1+\hat{x})(-1+q \hat{x})(q^2+a t^3 \hat{x})(q^2+a t^3 \hat{x}^2)(q^5+a t^3 \hat{x}^2)(1+a q^2 t^3 \hat{x}^2) (1+a q^3 t^3 \hat{x}^2)(-q^8-q^8 t \hat{x}-q^9 t \hat{x}+q^7 t^2 \hat{x}+q^8 t^2 \hat{x}-q^9 t^2 \hat{x}^2-a q^4 t^3 \hat{x}^2-a q^5 t^3 \hat{x}^2+q^7 t^3 \hat{x}^2+2 q^8 t^3 \hat{x}^2-a q^8 t^3 \hat{x}^2+q^9 t^3 \hat{x}^2-a q^9 t^3 \hat{x}^2-a q^6 t^4 \hat{x}^2-q^7 t^4 \hat{x}^2-a q^7 t^4 \hat{x}^2-a q^4 t^4 \hat{x}^3-a q^5 t^4 \hat{x}^3+a q^6 t^4 \hat{x}^3+2 a q^7 t^4 \hat{x}^3+a q^8 t^4 \hat{x}^3-a q^9 t^4 \hat{x}^3-a q^{10} t^4 \hat{x}^3+a q^3 t^5 \hat{x}^3+2 a q^4 t^5 \hat{x}^3-2 a q^6 t^5 \hat{x}^3-a q^7 t^5 \hat{x}^3+a q^8 t^5 \hat{x}^3+a q^9 t^5 \hat{x}^3+a q^5 t^6 \hat{x}^3+a q^6 t^6 \hat{x}^3+a q^7 t^6 \hat{x}^3+a q^6 t^5 \hat{x}^4+2 a q^7 t^5 \hat{x}^4+2 a q^8 t^5 \hat{x}^4+a q^9 t^5 \hat{x}^4-a^2 q t^6 \hat{x}^4+a q^3 t^6 \hat{x}^4+2 a q^4 t^6 \hat{x}^4-a^2 q^4 t^6 \hat{x}^4+a q^5 t^6 \hat{x}^4-2 a^2 q^5 t^6 \hat{x}^4-a q^6 t^6 \hat{x}^4-a^2 q^6 t^6 \hat{x}^4-a q^7 t^6 \hat{x}^4+a q^8 t^6 \hat{x}^4+2 a q^9 t^6 \hat{x}^4-a^2 q^9 t^6 \hat{x}^4+a q^{10} t^6 \hat{x}^4-a^2 q^2 t^7 \hat{x}^4-a q^3 t^7 \hat{x}^4-a^2 q^3 t^7 \hat{x}^4+a q^5 t^7 \hat{x}^4+a^2 q^5 t^7 \hat{x}^4+a q^6 t^7 \hat{x}^4-a^2 q^7 t^7 \hat{x}^4-a q^8 t^7 \hat{x}^4-a^2 q^8 t^7 \hat{x}^4+a^2 q^2 t^7 \hat{x}^5+2 a^2 q^3 t^7 \hat{x}^5+a^2 q^4 t^7 \hat{x}^5-a^2 q^5 t^7 \hat{x}^5-a^2 q^6 t^7 \hat{x}^5+a^2 q^7 t^7 \hat{x}^5+2 a^2 q^8 t^7 \hat{x}^5+a^2 q^9 t^7 \hat{x}^5+a^2 t^8 \hat{x}^5-a^2 q^2 t^8 \hat{x}^5+4 a^2 q^4 t^8 \hat{x}^5+5 a^2 q^5 t^8 \hat{x}^5+2 a^2 q^6 t^8 \hat{x}^5+a^2 q^9 t^8 \hat{x}^5+a^2 q t^9 \hat{x}^5+a^2 q^2 t^9 \hat{x}^5+a^2 q^3 t^9 \hat{x}^5+a^2 q^6 t^9 \hat{x}^5+a^2 q^7 t^9 \hat{x}^5+a^2 q^8 t^9 \hat{x}^5-a^2 q^5 t^8 \hat{x}^6-a^2 q^6 t^8 \hat{x}^6-a^2 q^7 t^8 \hat{x}^6-a^3 q t^9 \hat{x}^6-a^2 q^2 t^9 \hat{x}^6-a^3 q^2 t^9 \hat{x}^6-a^2 q^3 t^9 \hat{x}^6+a^2 q^4 t^9 \hat{x}^6+2 a^2 q^5 t^9 \hat{x}^6-a^3 q^5 t^9 \hat{x}^6+a^2 q^6 t^9 \hat{x}^6-a^3 q^6 t^9 \hat{x}^6-a^2 q^7 t^9 \hat{x}^6-a^2 q^8 t^9 \hat{x}^6+a^2 q t^{10} \hat{x}^6+a^3 q t^{10} \hat{x}^6+a^2 q^2 t^{10} \hat{x}^6-a^3 q^3 t^{10} \hat{x}^6-a^2 q^4 t^{10} \hat{x}^6-a^3 q^4 t^{10} \hat{x}^6+a^2 q^6 t^{10} \hat{x}^6+a^3 q^6 t^{10} \hat{x}^6+a^2 q^7 t^{10} \hat{x}^6+a^3 q^3 t^{10} \hat{x}^7+2 a^3 q^4 t^{10} \hat{x}^7+a^3 q^5 t^{10} \hat{x}^7+a^3 t^{11} \hat{x}^7+2 a^3 q t^{11} \hat{x}^7-a^3 q^3 t^{11} \hat{x}^7-a^3 q^4 t^{11} \hat{x}^7+2 a^3 q^5 t^{11} \hat{x}^7+a^3 q^6 t^{11} \hat{x}^7+a^3 q^2 t^{12} \hat{x}^7+a^3 q^3 t^{12} \hat{x}^7+a^3 q^4 t^{12} \hat{x}^7-a^4 q^2 t^{12} \hat{x}^8-a^3 q^3 t^{12} \hat{x}^8-a^3 q^4 t^{12} \hat{x}^8+a^3 q^2 t^{13} \hat{x}^8+a^4 q^2 t^{13} \hat{x}^8+a^3 q^3 t^{13} \hat{x}^8+a^4 q t^{14} \hat{x}^9+a^4 q^2 t^{14} \hat{x}^9) \hat{y}^2$\\[3pt] \rule{0pt}{2mm}
&\footnotesize$+a^3 q^{-5} t^6 (-1+\hat{x})\hat{x}^2(-1+q \hat{x})(-1+q^2 \hat{x})(q^2+a t^3 \hat{x}^2)(1+a q^3 t^3 \hat{x}^2)(1+a q^4 t^3 \hat{x}^2)(1+a q^5 t^3 \hat{x}^2)(-q^4-q^5-q^5 t \hat{x}-q^6 t \hat{x}+q^4 t^2 \hat{x}+q^5 t^2 \hat{x}-a q^3 t^3 \hat{x}-a q t^3 \hat{x}^2-a q^2 t^3 \hat{x}^2-a q^6 t^3 \hat{x}^2-a q^7 t^3 \hat{x}^2-2 a q^3 t^4 \hat{x}^2-a q^4 t^4 \hat{x}^2-a q^5 t^4 \hat{x}^2+a q^3 t^5 \hat{x}^2+a q^4 t^4 \hat{x}^3+a q^5 t^4 \hat{x}^3+a q^6 t^4 \hat{x}^3+a q t^5 \hat{x}^3+a q^2 t^5 \hat{x}^3+a q^6 t^5 \hat{x}^3+a q^7 t^5 \hat{x}^3-a^2 t^6 \hat{x}^3+a q^3 t^6 \hat{x}^3-a^2 q^5 t^6 \hat{x}^3-a^2 q^3 t^6 \hat{x}^4-a^2 q^4 t^6 \hat{x}^4-a^2 t^7 \hat{x}^4+a^2 q^3 t^7 \hat{x}^4+a^2 q^4 t^7 \hat{x}^4-a^2 q^5 t^7 \hat{x}^4+a^2 t^8 \hat{x}^4+a^2 q^5 t^8 \hat{x}^4+a^2 q^3 t^8 \hat{x}^5+a^2 q^4 t^8 \hat{x}^5+a^2 t^9 \hat{x}^5-a^3 q^2 t^9 \hat{x}^5+a^2 q^5 t^9 \hat{x}^5-a^3 q^2 t^{10} \hat{x}^6+a^3 q^2 t^{11} \hat{x}^6+a^3 q^2 t^{12} \hat{x}^7) \hat{y}$\\[3pt] \rule{0pt}{2mm}
&\footnotesize$-a^4 q^4 t^{10}(-1+\hat{x}) \hat{x}^4(-1+q \hat{x})(-1+q^2 \hat{x})(-1+q^3 \hat{x})(1+a q^4 t^3 \hat{x}^2)(1+a q^5 t^3 \hat{x}^2)(1+a q^6 t^3 \hat{x}^2)(1+a q^7 t^3 \hat{x}^2)$
\tabularnewline
\hline 
 \end{tabular}
\caption{Quantum super-$A$-polynomial of the knots ${\bf 6_1}$}
\end{center}
\label{tab:qapoly61}
\end{table}

\begin{table}[H]
\renewcommand{\arraystretch}{1}
\begin{center}
\begin{tabular}{|c|p{12cm}|}
\hline 
\textbf{ Knot} & $A^{\rm{super}}(K;x=-\mu,y=\frac{1+\mu}{1+U\mu}\lambda;a=U,t=-1)$
\tabularnewline\hline 
\hline \rule{0pt}{5mm}
${\bf 3_2}$ &$\frac{1+\mu}{(1+U\mu)^2}$\footnotesize{$(\lambda ^2 \mu +\lambda ^2-\lambda  \mu ^4 U^3+\mu ^4 U^3-\lambda  \mu ^3 U^2+2 \lambda  \mu ^2 U^2+\mu ^3 U^2-2 \lambda  \mu ^2 U-\lambda  \mu  U-\lambda  U)$}
\tabularnewline
\hline \rule{0pt}{5mm}
${\bf 4_1}$ &$\frac{(1+\mu)^2}{U{\mu}^2(1+U\mu)^3}$\footnotesize{$(U \lambda - \lambda^2 + 2 U \lambda \mu - 
 2 U \lambda^2 \mu - U^2 mu^2 + U \lambda^3 \mu^2 - 
 U^3 \mu^3 + U \lambda^3 \mu^3 + 2 U^3 \lambda \mu^4 - 
 2 U^2 \lambda^2 \mu^4 + U^3 \lambda \mu^5 - 
 U^3 \lambda^2 \mu^5)$}
\tabularnewline
\hline \rule{0pt}{5mm}
${\bf 5_2}$ &$\frac{(1+\mu)^3}{(1+U\mu)^4}$\footnotesize{$(U^2 \lambda ^2-2 U \lambda ^3+\lambda ^4+3 U^2 \lambda ^2 \mu -4 U \lambda ^3 \mu +\lambda ^4 \mu -U^3 \lambda  \mu ^2+5 U^2 \lambda ^2 \mu ^2-4 U^3 \lambda ^2 \mu ^2-3 U \lambda ^3 \mu ^2+3 U^2 \lambda ^3 \mu ^2-2 U^3 \lambda  \mu ^3+3 U^2 \lambda ^2 \mu ^3-3 U^3 \lambda ^2 \mu ^3+2 U^2 \lambda ^3 \mu ^3-U^3 \lambda  \mu ^4-4 U^3 \lambda ^2 \mu ^4+6 U^4 \lambda ^2 \mu ^4-U^2 \lambda ^3 \mu ^4+2 U^4 \lambda  \mu ^5+3 U^3 \lambda ^2 \mu ^5-3 U^4 \lambda ^2 \mu ^5-2 U^3 \lambda ^3 \mu ^5-3 U^4 \lambda  \mu ^6+3 U^5 \lambda  \mu ^6+5 U^4 \lambda ^2 \mu ^6-4 U^5 \lambda ^2 \mu ^6-U^4 \lambda ^3 \mu ^6+U^5 \mu ^7-4 U^5 \lambda  \mu ^7+3 U^5 \lambda ^2 \mu ^7+U^6 \mu ^8-2 U^6 \lambda  \mu ^8+U^6 \lambda ^2 \mu ^8)$}
\tabularnewline
\hline \rule{0pt}{5mm}
${\bf 6_1}$ & $\frac{(1+\mu)^4}{U^2{\mu}^4(1+U\mu)^5}$ \footnotesize{$(-U^2 \lambda ^2+2 U \lambda ^3-\lambda ^4-4 U^2 \lambda ^2 \mu +4 U \lambda ^3 \mu +2 U^2 \lambda ^3 \mu -2 U \lambda ^4 \mu +2 U^3 \lambda  \mu ^2-6 U^2 \lambda ^2 \mu ^2+2 U^3 \lambda ^2 \mu ^2+3 U^2 \lambda ^3 \mu ^2-U^2 \lambda ^4 \mu ^2+4 U^3 \lambda  \mu ^3+U^4 \lambda  \mu ^3-U^3 \lambda ^2 \mu ^3-4 U^3 \lambda ^3 \mu ^3-U^4 \mu ^4+U^4 \lambda  \mu ^4-3 U^4 \lambda ^2 \mu ^4+2 U^3 \lambda ^4 \mu ^4+U^2 \lambda ^5 \mu ^4-U^5 \mu ^5-2 U^5 \lambda  \mu ^5+3 U^4 \lambda ^3 \mu ^5-U^3 \lambda ^4 \mu ^5+U^2 \lambda ^5 \mu ^5+4 U^5 \lambda ^2 \mu ^6+U^4 \lambda ^3 \mu ^6-4 U^3 \lambda ^4 \mu ^6-U^4 \lambda ^4 \mu ^6+U^6 \lambda  \mu ^7-3 U^5 \lambda ^2 \mu ^7+6 U^4 \lambda ^3 \mu ^7-2 U^5 \lambda ^3 \mu ^7-2 U^4 \lambda ^4 \mu ^7+2 U^6 \lambda  \mu ^8-4 U^5 \lambda ^2 \mu ^8-2 U^6 \lambda ^2 \mu ^8+4 U^5 \lambda ^3 \mu ^8+U^6 \lambda  \mu ^9-2 U^6 \lambda ^2 \mu ^9+U^6 \lambda ^3 \mu ^9)$}
\tabularnewline
\hline \rule{0pt}{5mm}
${\bf 7_2}$ &$-\frac{(1+\mu)^5}{(1+U\mu)^6}$\footnotesize{$(U^3 \lambda ^3-3 U^2 \lambda ^4+3 U \lambda ^5-\lambda ^6+5 U^3 \lambda ^3 \mu -11 U^2 \lambda ^4 \mu +7 U \lambda ^5 \mu -\lambda ^6 \mu -2 U^4 \lambda ^2 \mu ^2+14 U^3 \lambda ^3 \mu ^2-6 U^4 \lambda ^3 \mu ^2-17 U^2 \lambda ^4 \mu ^2+10 U^3 \lambda ^4 \mu ^2+5 U \lambda ^5 \mu ^2-4 U^2 \lambda ^5 \mu ^2-7 U^4 \lambda ^2 \mu ^3+19 U^3 \lambda ^3 \mu ^3-15 U^4 \lambda ^3 \mu ^3-10 U^2 \lambda ^4 \mu ^3+16 U^3 \lambda ^4 \mu ^3-3 U^2 \lambda ^5 \mu ^3+U^5 \lambda  \mu ^4-10 U^4 \lambda ^2 \mu ^4+5 U^5 \lambda ^2 \mu ^4+10 U^3 \lambda ^3 \mu ^4-20 U^4 \lambda ^3 \mu ^4+15 U^5 \lambda ^3 \mu ^4+9 U^3 \lambda ^4 \mu ^4-10 U^4 \lambda ^4 \mu ^4+2 U^5 \lambda  \mu ^5-5 U^4 \lambda ^2 \mu ^5+8 U^5 \lambda ^2 \mu ^5-9 U^4 \lambda ^3 \mu ^5+10 U^5 \lambda ^3 \mu ^5-6 U^4 \lambda ^4 \mu ^5+U^5 \lambda  \mu ^6+3 U^5 \lambda ^2 \mu ^6+12 U^5 \lambda ^3 \mu ^6-20 U^6 \lambda ^3 \mu ^6+3 U^4 \lambda ^4 \mu ^6+U^3 \lambda ^5 \mu ^6-6 U^6 \lambda ^2 \mu ^7-9 U^5 \lambda ^3 \mu ^7+10 U^6 \lambda ^3 \mu ^7-5 U^4 \lambda ^4 \mu ^7+8 U^5 \lambda ^4 \mu ^7+2 U^4 \lambda ^5 \mu ^7+9 U^6 \lambda ^2 \mu ^8-10 U^7 \lambda ^2 \mu ^8+10 U^5 \lambda ^3 \mu ^8-20 U^6 \lambda ^3 \mu ^8+15 U^7 \lambda ^3 \mu ^8-10 U^5 \lambda ^4 \mu ^8+5 U^6 \lambda ^4 \mu ^8+U^5 \lambda ^5 \mu ^8-3 U^7 \lambda  \mu ^9-10 U^6 \lambda ^2 \mu ^9+16 U^7 \lambda ^2 \mu ^9+19 U^6 \lambda ^3 \mu ^9-15 U^7 \lambda ^3 \mu ^9-7 U^6 \lambda ^4 \mu ^9+5 U^7 \lambda  \mu ^{10}-4 U^8 \lambda  \mu ^{10}-17 U^7 \lambda ^2 \mu ^{10}+10 U^8 \lambda ^2 \mu ^{10}+14 U^7 \lambda ^3 \mu ^{10}-6 U^8 \lambda ^3 \mu ^{10}-2 U^7 \lambda ^4 \mu ^{10}-U^8 \mu ^{11}+7 U^8 \lambda  \mu ^{11}-11 U^8 \lambda ^2 \mu ^{11}+5 U^8 \lambda ^3 \mu ^{11}-U^9 \mu ^{12}+3 U^9 \lambda  \mu ^{12}-3 U^9 \lambda ^2 \mu ^{12}+U^9 \lambda ^3 \mu ^{12})$} 
\tabularnewline
\hline \rule{0pt}{5mm}
${\bf 8_1}$ &$\frac{(1+\mu)^6}{U^3{\mu}^6(1+U\mu)^7}$\footnotesize{$(U^3 \lambda ^3-3 U^2 \lambda ^4+3 U \lambda ^5-\lambda ^6+6 U^3 \lambda ^3 \mu -12 U^2 \lambda ^4 \mu -2 U^3 \lambda ^4 \mu +6 U \lambda ^5 \mu +4 U^2 \lambda ^5 \mu -2 U \lambda ^6 \mu -3 U^4 \lambda ^2 \mu ^2+18 U^3 \lambda ^3 \mu ^2-4 U^4 \lambda ^3 \mu ^2-15 U^2 \lambda ^4 \mu ^2-2 U^3 \lambda ^4 \mu ^2+6 U^2 \lambda ^5 \mu ^2+U^3 \lambda ^5 \mu ^2-U^2 \lambda ^6 \mu ^2-12 U^4 \lambda ^2 \mu ^3-U^5 \lambda ^2 \mu ^3+20 U^3 \lambda ^3 \mu ^3-6 U^4 \lambda ^3 \mu ^3-5 U^3 \lambda ^4 \mu ^3+8 U^4 \lambda ^4 \mu ^3-4 U^3 \lambda ^5 \mu ^3+3 U^5 \lambda  \mu ^4-15 U^4 \lambda ^2 \mu ^4+2 U^5 \lambda ^2 \mu ^4+8 U^5 \lambda ^3 \mu ^4+6 U^4 \lambda ^4 \mu ^4-4 U^4 \lambda ^5 \mu ^4+6 U^5 \lambda  \mu ^5+2 U^6 \lambda  \mu ^5+4 U^6 \lambda ^2 \mu ^5+U^5 \lambda ^3 \mu ^5-13 U^5 \lambda ^4 \mu ^5-U^6 \mu ^6+2 U^6 \lambda  \mu ^6+U^6 \lambda ^2 \mu ^6-12 U^6 \lambda ^3 \mu ^6+6 U^5 \lambda ^5 \mu ^6+3 U^4 \lambda ^6 \mu ^6+U^3 \lambda ^7 \mu ^6-U^7 \mu ^7-3 U^7 \lambda  \mu ^7-6 U^7 \lambda ^2 \mu ^7+12 U^6 \lambda ^4 \mu ^7-U^5 \lambda ^5 \mu ^7-2 U^4 \lambda ^6 \mu ^7+U^3 \lambda ^7 \mu ^7+13 U^7 \lambda ^3 \mu ^8-U^6 \lambda ^4 \mu ^8-4 U^6 \lambda ^5 \mu ^8-6 U^4 \lambda ^6 \mu ^8-2 U^5 \lambda ^6 \mu ^8+4 U^8 \lambda ^2 \mu ^9-6 U^7 \lambda ^3 \mu ^9-8 U^7 \lambda ^4 \mu ^9+15 U^5 \lambda ^5 \mu ^9-2 U^6 \lambda ^5 \mu ^9-3 U^5 \lambda ^6 \mu ^9+4 U^8 \lambda ^2 \mu ^{10}+5 U^7 \lambda ^3 \mu ^{10}-8 U^8 \lambda ^3 \mu ^{10}-20 U^6 \lambda ^4 \mu ^{10}+6 U^7 \lambda ^4 \mu ^{10}+12 U^6 \lambda ^5 \mu ^{10}+U^7 \lambda ^5 \mu ^{10}+U^9 \lambda  \mu ^{11}-6 U^8 \lambda ^2 \mu ^{11}-U^9 \lambda ^2 \mu ^{11}+15 U^7 \lambda ^3 \mu ^{11}+2 U^8 \lambda ^3 \mu ^{11}-18 U^7 \lambda ^4 \mu ^{11}+4 U^8 \lambda ^4 \mu ^{11}+3 U^7 \lambda ^5 \mu ^{11}+2 U^9 \lambda  \mu ^{12}-6 U^8 \lambda ^2 \mu ^{12}-4 U^9 \lambda ^2 \mu ^{12}+12 U^8 \lambda ^3 \mu ^{12}+2 U^9 \lambda ^3 \mu ^{12}-6 U^8 \lambda ^4 \mu ^{12}+U^9 \lambda  \mu ^{13}-3 U^9 \lambda ^2 \mu ^{13}+3 U^9 \lambda ^3 \mu ^{13}-U^9 \lambda ^4 \mu ^{13})$}
\tabularnewline
\hline
\end{tabular}\caption{Classical super polynomials at the point  $x=-\mu,\, y=\frac{1+\mu}{1+U\mu}\lambda,\, a=U,\,t=-1$. The second factors are equal to the augmentation polynomials obtained by Lenny Ng \cite{Ng:2012} up to sign.}
\label{tab:augmented}
\end{center}
\end{table}
\pagebreak

\bibliography{CS}{}
\bibliographystyle{JHEP}
\end{document}